\documentclass[letterpaper]{article}

\usepackage[T1]{fontenc}

\usepackage{geometry}
\usepackage{setspace}
\usepackage[autostyle=false]{csquotes}

\usepackage[
    style=chem-acs,
    articletitle=true,
    autocite=superscript,  % \autocite → superscript; no need to alias \autocite
    maxbibnames=20,
    minbibnames=20,
    backend=biber
]{biblatex}

\AtEveryBibitem{%
  \ifentrytype{article}{%
    \clearfield{number}%
    \clearfield{issue}%
  }{}%
}

\usepackage{graphics}
\usepackage{subcaption}
\usepackage{hyperref}
\usepackage{comment}

\usepackage{float}
\newfloat{scheme}{htbp}{los}
\floatname{scheme}{Scheme}
\floatname{chart}{Chart}
\newfloat{graph}{htbp}{loh}

\usepackage{chemformula} % Formulas using \ch{}
\usepackage[version = 4]{mhchem} % Formulas using \ce{}
\usepackage{amsmath}
\usepackage{amssymb}

\usepackage{authblk}
\author[1]{Ayman Hussein}
\author[1,2,3,4]{Ralf Bundschuh}
\affil[1]{Department of Physics, The Ohio State University, Columbus, OH 43210, United States}
\affil[2]{Department of Chemistry and Biochemistry, The Ohio State University, Columbus, OH 43210, United States}
\affil[3]{Division of Hematology, Department of Internal Medicine, The Ohio State University, Columbus, OH 43210, United States} 
\affil[4]{Center for RNA Biology, The Ohio State University, Columbus, OH 43210, United States}

\title{A kinetic model of shear-induced rupture of short dsDNA}
\date{*Email: bundschuh.2@osu.edu}

\begin{document}

\maketitle

\begin{abstract}
Force-induced dissociation of short double-stranded DNA (dsDNA) is central to single-molecule biophysics and DNA nanotechnology, yet a physically grounded kinetic description of shear-induced rupture for finite-length constructs remains lacking. Here we develop a master equation framework built on a force-dependent nucleation-zipper pathway with single-base transitions, enabling direct calculation of dissociation rates and transition state distances over a broad force range. 
Applied to a DNA-gold nanoparticle-DNA construct under constant shear force, the model accurately reproduces the experimental room-temperature data in the covered force regime and provides a unified interpretation of prior measurements on similarly sheared duplexes across all force regimes. 
A central result is that the three-dimensional helical geometry of dsDNA is essential for correctly defining the end to end distance under shear in the rod-like polymer model of short dsDNA. We further show that the extracted transition state distances are robust to variations in ssDNA polymer parameters within the experimentally relevant regime. 
Finally, we analyze the temperature dependence of the transition state distance and discuss how our framework captures globally-heated rupture while identifying the additional complications introduced by localized plasmonic heating in gold nanoparticle-coupled constructs. These results provide a predictive kinetic foundation for interpreting force-rupture experiments and for designing force- and temperature-actuated DNA nanostructures.
\end{abstract}

\section*{Keywords}
force-induced rupture of short dsDNA, transition state distance, master equation, DNA helical geometry, polymer models of ssDNA and dsDNA

Double-stranded DNA (dsDNA) dissociation under mechanical load is central to both molecular biophysics and DNA nanotechnology. In biological contexts, forces acting on nucleic acids accompany replication, transcription, recombination, and repair, while in engineered systems, short DNA duplexes serve as programmable mechanical elements in nanostructures and force sensors~\autocite{Vologodskii_2015, origami_review_dynamic_2020}. 
Force-rupture of short duplexes shows a subtle interplay between duplex thermodynamics, the elasticity of the force-bearing single-stranded and double-stranded segments, the geometry of force application, and the finite timescale of observation~\autocite{oxdna_force-induced_2015}. This complexity is especially pronounced in the shear geometry, where antiparallel forces act on opposite strands and the molecule must traverse a force-dependent free energy barrier before full dissociation. \\
Single-molecule force spectroscopy provides a direct route to address this problem, linking externally applied force to molecular lifetimes and rupture trajectories. General force-spectroscopy frameworks have demonstrated that intrinsic off-rates, transition state locations, and activation free energies can all be extracted from mechanical rupture data~\autocite{bells, schumakovitch_temperature_2002, whitley_elasticity_2017, kabtiyal_localized_2024}. Yet the physical interpretation of these quantities depends critically on the validity of the assumed reaction coordinate and on how the mechanical response of the molecule and its linkers is modeled~\autocite{dudko_intrinsic_2006, dudko_theory_2008}. For DNA in particular, the nearest-neighbor thermodynamic framework and nucleation-zipper kinetic models describe bulk hybridization well, but their translation to force-dependent rupture of finite-length constructs remains incomplete~\autocite{ashwood_kinetics_2025}.\\
Here, we develop a kinetic model of shear-force-induced rupture of short dsDNA applied to the experiment presented in Kabtiyal et al.~\autocite{kabtiyal_localized_2024}, formulated as a master equation system~\autocite{master_cosmic_1940,hanggi_reaction-rate_1990,vankampen2007spp} and built on a force-dependent nucleation-zipper pathway~\autocite{nuc_zip_1971,ashwood_kinetics_2025} with single-base transitions. We show that the model accurately reproduces the experimental room-temperature data, and provides a unified interpretation of prior measurements~\autocite{schumakovitch_temperature_2002,whitley_elasticity_2017,dna_bow_weak_2023} on similarly sheared duplexes across all force regimes. 
A central result is that the three-dimensional helical geometry of dsDNA~\autocite{Vologodskii_2015} is essential for correctly defining the mechanical response under shear at this length scale. We further examine the role of the force-extension description assigned to the released ssDNA segments, showing that the extracted transition state distances are insensitive to the behavior of very short strand lengths and robust across reasonable estimates of ssDNA polymer parameters~\autocite{andersen_stretching_2022, bosco_elastic_2014, whitley_elasticity_2017}. Finally, we analyze the temperature dependence of the transition state distance and discuss the additional care required when interpreting plasmonically excited rupture data, where localized heating and position-dependent temperature profiles introduce complications absent in the constant-temperature case~\autocite{jain_calculated_2006, amendola_surface_2017}. 
Together, these results establish a physically grounded kinetic framework for force-induced DNA rupture with direct relevance to force- and temperature-actuated DNA nanostructures.

%%%%%%%%%%%%%%%%%%%%%%%%%%%%%%%%%%%%%%%%%%%%%%%%%%%%%%%
\newpage
\section{Results and discussion}
%%%%%%%%%%%
\subsection{Review of the experimental setup}
\begin{figure}[ht!]
    \centering
    \includegraphics[width=\textwidth]{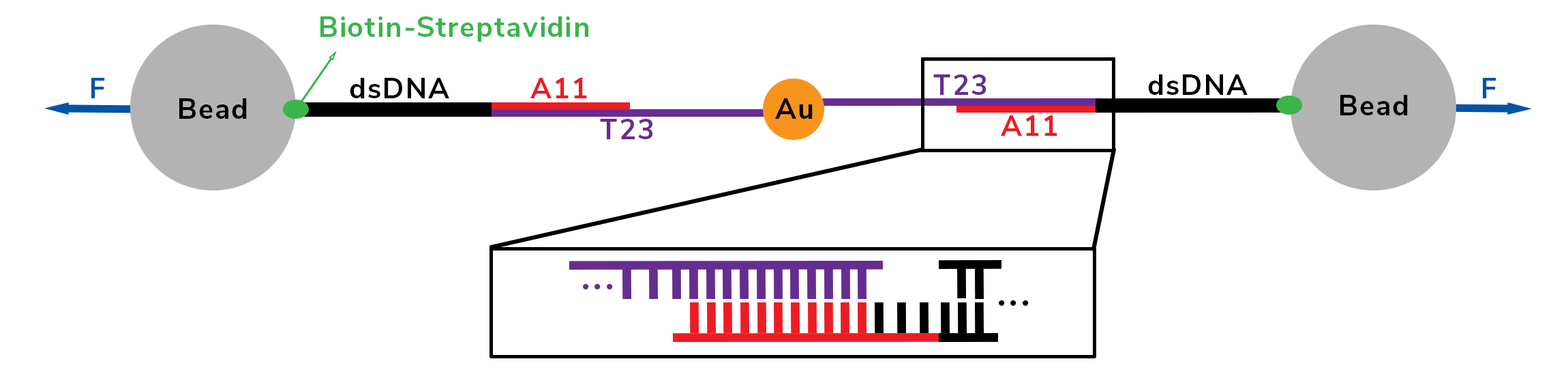}
    \caption{Schematic of the bead-DNA-AuNP-DNA-bead setup in a dual-trap under constant shear force (F). Zoomed out: Schematic of the binding of A$11$/T$11$ with four-nucleotide spacing followed by dsDNA handles.}
    \label{fig:AuNP}
\end{figure}

Before introducing our model of shear-force rupture of dsDNA, we review the experimental setup in Kabtiyal et al.~\autocite{kabtiyal_localized_2024} that we base our modeling on.
The experimental setup consists of a $15$nm central gold nano-particle (AuNP) functionalized with ssDNA strands consisting of $23$ thymine bases (T$23$) each, two of which hybridize to single-stranded stretches of $11$ adenine bases (A$11$). The $3^\prime$ ends of the polyA, stretches extended with dsDNA tethers, are attached to two beads held in a Lumicks C-Trap system using biotin-streptavidin attachments leading to the final construct in Fig.~\ref{fig:AuNP}. Once the construct is formed, a single-tether of dsDNA-AuNP-dsDNA is confirmed through force-extension comparison with the Marko-Siggia force-extension formula~\autocite{marko_stretching_1995}. Then, a constant force is applied to the bead-dsDNA-AuNP-dsDNA-bead construct in a buffer channel that has $0.5\times$TE, $100$~mM \ce{NaCl}, $5.5$~mM \ce{MgCl2}, and $0.05$\%~Tween $20$.\\ 
Within this setup, the short dsDNA sequence ($11$bp of polyA-polyT) is subject to constant shear-force ($2.5, 5, 10$ pN) from the two ends ($3'-3'$) until rupture is observed and the rupture time is recorded. This is repeated with (excited) and without (dark) localized laser heating of the AuNP for $20$ instances per force, the rupture times are recorded, and the instances that do not rupture within a $600$s window are denoted as unruptured (U). We will focus on the dark rupture for the bulk of the manuscript, but we will return to the discussion of the excited rupture at the end.

\subsection{Data analysis affects the inferred transition state distance}
\begin{figure}[ht!]
    \centering
    \includegraphics[width=0.5\textwidth]{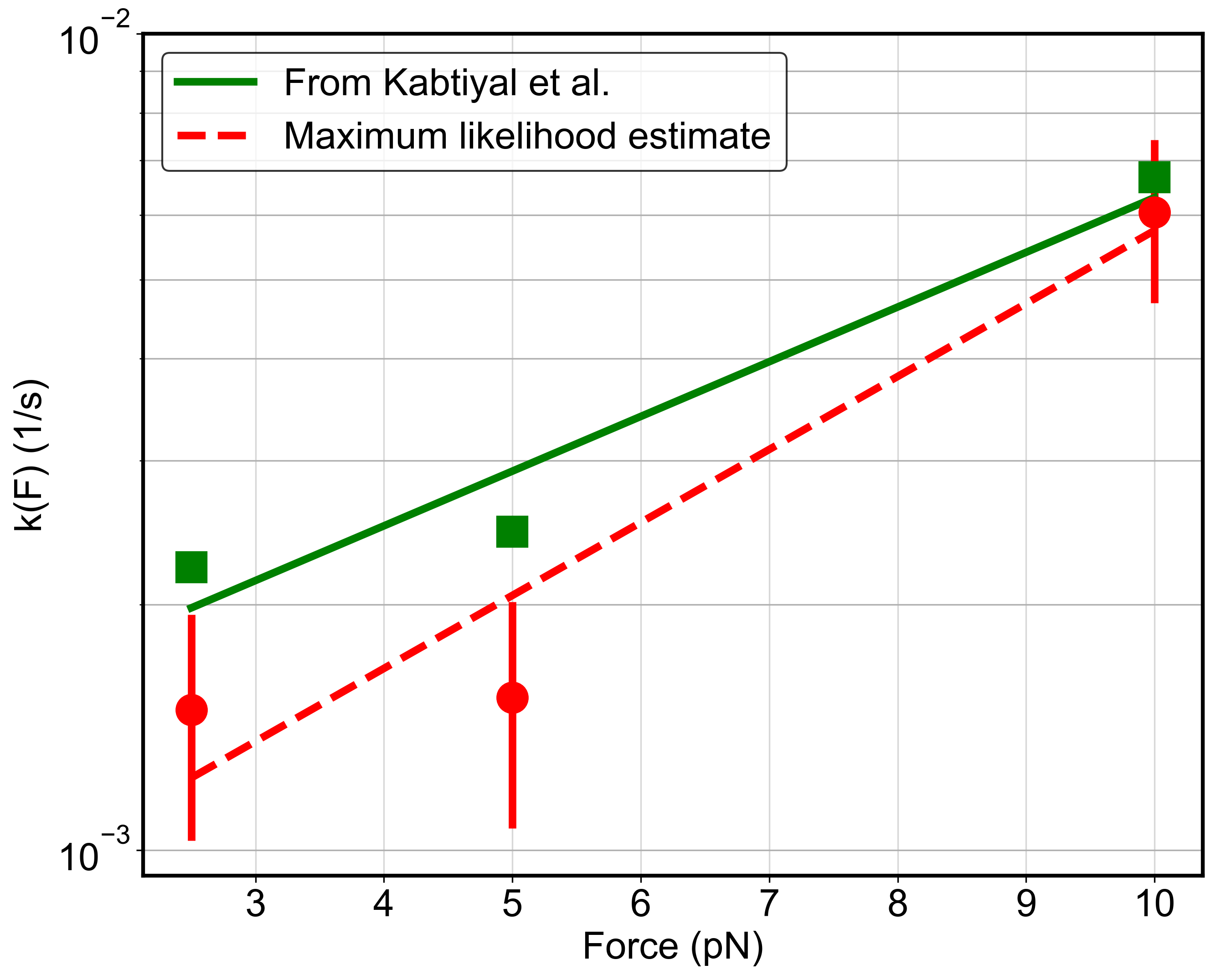}
    \caption{A comparison of the different analyses of the experimental \textit{dark-rupture} times from~\autocite{kabtiyal_localized_2024}, and their corresponding Bell's linear fits (Eq.~\ref{equ:bell}). Circles and dashed line; Maximum-likelihood estimate (MLE), squares and solid line; original data of~\autocite{kabtiyal_localized_2024}.}
    \label{fig:lnkvsf-exp-dark}
\end{figure}

There are two main ways to analyze the recorded rupture times in order to obtain reaction off-rates. On the one hand, one collects extensive data and fits a single exponential distribution to the cumulative probabilities ($P_c(t)$) obtained from the rupture times. On the other hand, when limited data is available, a maximum-likelihood estimate (MLE) of the reaction off-rates is better suited~\autocite{sm_theory_book, lawless_statistical_2003}, see \textit{Supporting Information} for derivation.\\ 
In the original analysis of Kabtiyal et al.~\autocite{kabtiyal_localized_2024}, the right-censored rupture times ($<600$s) were first binned using a bin width of 30s before computing the cumulative distributions. A single-exponential function $(P_c(t)\approx1 - e^{-k(F) t})$ was then fitted to those binned cumulative distributions for each applied force ($F$), where $k(F)$ is the dissociation rate at that force. 
However, given the limited data for the dark rupture at $2.5$pN and $5$pN, where only about half the instances ruptured, a reanalysis using the MLE method is essential to analyze the dark-rupture data of Kabtiyal et al.~\autocite{kabtiyal_localized_2024}.\\
Once the reaction off-rates are calculated for the different forces, the transition state distance is obtained by invoking Bell's linear-fit formula \autocite{bells}
\begin{align}
    \ln k(F)=\ln k_a-\frac{\Delta G_{\text{duplex}}}{k_BT}+\frac{F\cdot d}{k_BT},
    \label{equ:bell}
\end{align}
where $\Delta G_{\text{duplex}}$ is the duplex dissociation free energy, $d$ is the transition state distance, $k_a$ is the attempt rate, $k_B$ is Boltzmann's constant, and $T$ is the temperature in Kelvin.\\
A comparison of the MLE obtained off-rates and the reported values from Kabtiyal et al.~\autocite{kabtiyal_localized_2024} for the dark rupture case is shown in Fig.~\ref{fig:lnkvsf-exp-dark} along with the corresponding Bell's linear fits.
With this MLE reanalysis, we observe a larger transition state distance of $d^\text{Exp-MLE}_{\text{Dark}}=0.85\pm0.20$nm, compared to the reported $d^\text{Exp}_{\text{Dark}}=0.60\pm0.16$nm.

\subsection{A kinetic model of force-induced shear-rupture of short dsDNA}
\begin{figure}[ht!]
    \centering
    \includegraphics[width=0.5\textwidth]{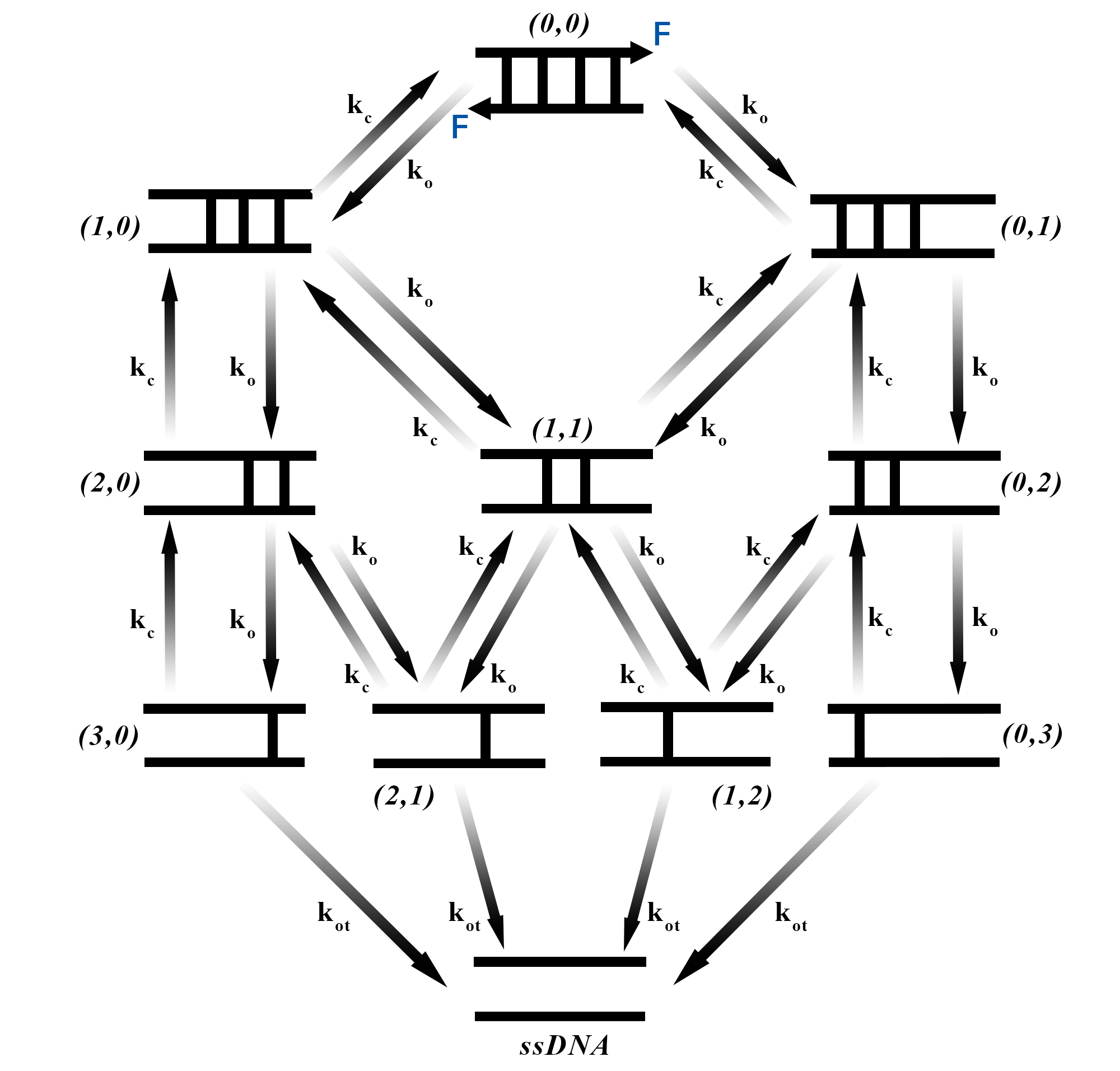}
    \caption{Demonstration of the duplex rupture under shear-force (F) for a 4bp duplex. The numbers in $(\cdot,\cdot)$ count the ruptured base pairs from each side. For simplicity,
    the transition opening ($k_o$), closing ($k_c$), and opening terminal ($k_{ot}$) rates are shown without their associated dependencies on the thermodynamic and/or mechanical properties of the transitioning states (see the \textit{Methods Section} for detailed definition of the transition rates). Similarly, 
    the force (F) is explicitly shown only for the initial duplex state while it is implied for the other states.}
    \label{fig:model-tree}
\end{figure}

We model force-induced shear-rupture \textit{via} a force-dependent version of the nucleation-zipper kinetic model~\autocite{nuc_zip_1971,ashwood_kinetics_2025}. As illustrated in Fig.~\ref{fig:model-tree} for a duplex of $N_{\text{\text{bp}}}=4$, the states are labeled by the ordered pair $(i,j)$, where $i\ge0$ and $j\ge0$ denote the number of ruptured base pairs from the left and right end, respectively, such that $i+j\le N_{\text{\text{bp}}}-1$. Under these conventions, the total number of states (system size) is given by $N_\text{states} = 1 + \frac{1}{2}N_\text{bp}(N_\text{bp}+1)$, which yields 11 states for the 4bp example shown in Fig.~\ref{fig:model-tree}, and 67 states for the 11bp duplex studied by Kabtiyal et al.~\autocite{kabtiyal_localized_2024} (see \textit{Methods Section} for a detailed description of the system of master equations).
\\
The system starts from the fully hybridized duplex state of $N_{\text{\text{bp}}}$ paired bases and the time evolution of the probability of the ruptured single-stranded (ssDNA) state is computed through a well-defined network of single-base opening or closing events, with opening ($k_o$), closing ($k_c$), and terminal opening ($k_{ot}$) rates, occurring independently from either the left or the right end. 
It is this probability of the ruptured state at time $t$ that corresponds to the cumulative probability ($P_c(t)$) of the rupture times obtained in the experiment, and it is thus the quantity we use to extract off-rates through single-exponential fitting (see the \textit{Methods Section} for detailed evaluation of the cumulative probability). 
In this formulation, we neglect the formation of possible bulges, internal loops, and sliding as they cost more than double the free energy of the aforementioned base pair breakage at the ends, rendering them less probable. 
Below, we also show how this simplified model captures the experimental behavior with sufficient accuracy without the need to introduce additional complexity.\\
Finally, we emphasize that the transition rates are generally different and force-dependent, but significant simplifications arise for homoduplex sequences (e.g., polyA/polyT) at constant temperature, as in the dark rupture experiment of~\autocite{kabtiyal_localized_2024} which makes it the perfect platform to test our model (see \textit{Methods Section} for detailed definitions of the transition rates). The force-dependence arises from the force-extension response of short ssDNA and dsDNA. For dsDNA, we use a rod-like polymer model with helical geometry dependent rod length. For ssDNA, a global interpolation formula (GIF) of the worm-like-chain model with experimentally-determined persistence length ($\lambda_\text{ss}$) and interphosphate distance ($l_\text{ss}$) best captures the ssDNA force-extension response at the short length scales relevant in this experiment~\autocite{andersen_stretching_2022,bosco_elastic_2014}. The choice of the different polymer models for ssDNA and dsDNA is discussed further below.

\subsection{The model accurately captures experimental dark rupture at all force regimes}
\begin{figure}[ht!]
    \centering
    \begin{subfigure}[ht!]{0.5\textwidth}
        \centering
        \includegraphics[width=\textwidth]{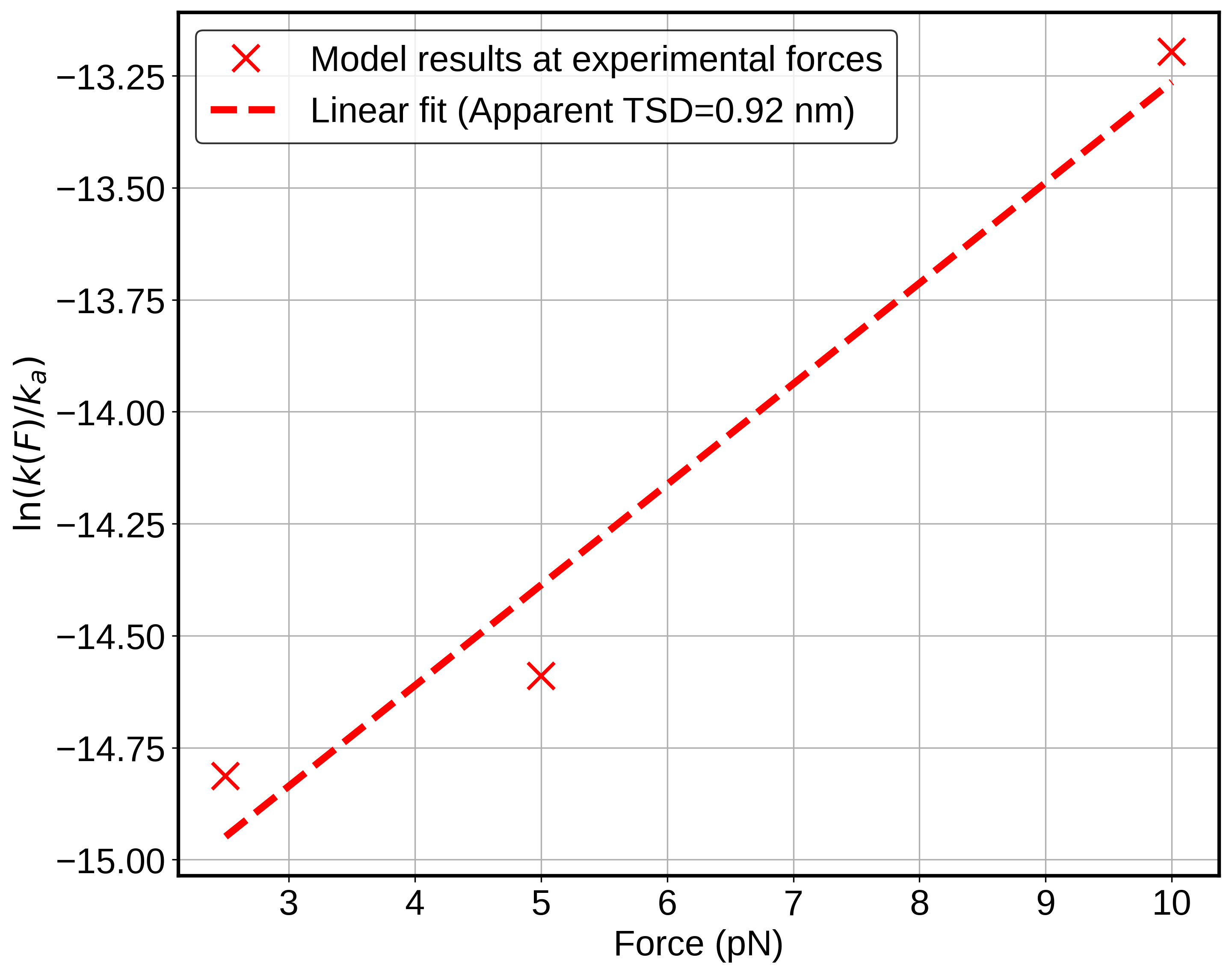}
        \caption{}
        \label{fig:MM-T25-Three}
    \end{subfigure}%
    ~ 
    \begin{subfigure}[ht!]{0.5\textwidth}
        \centering
        \includegraphics[width=\textwidth]{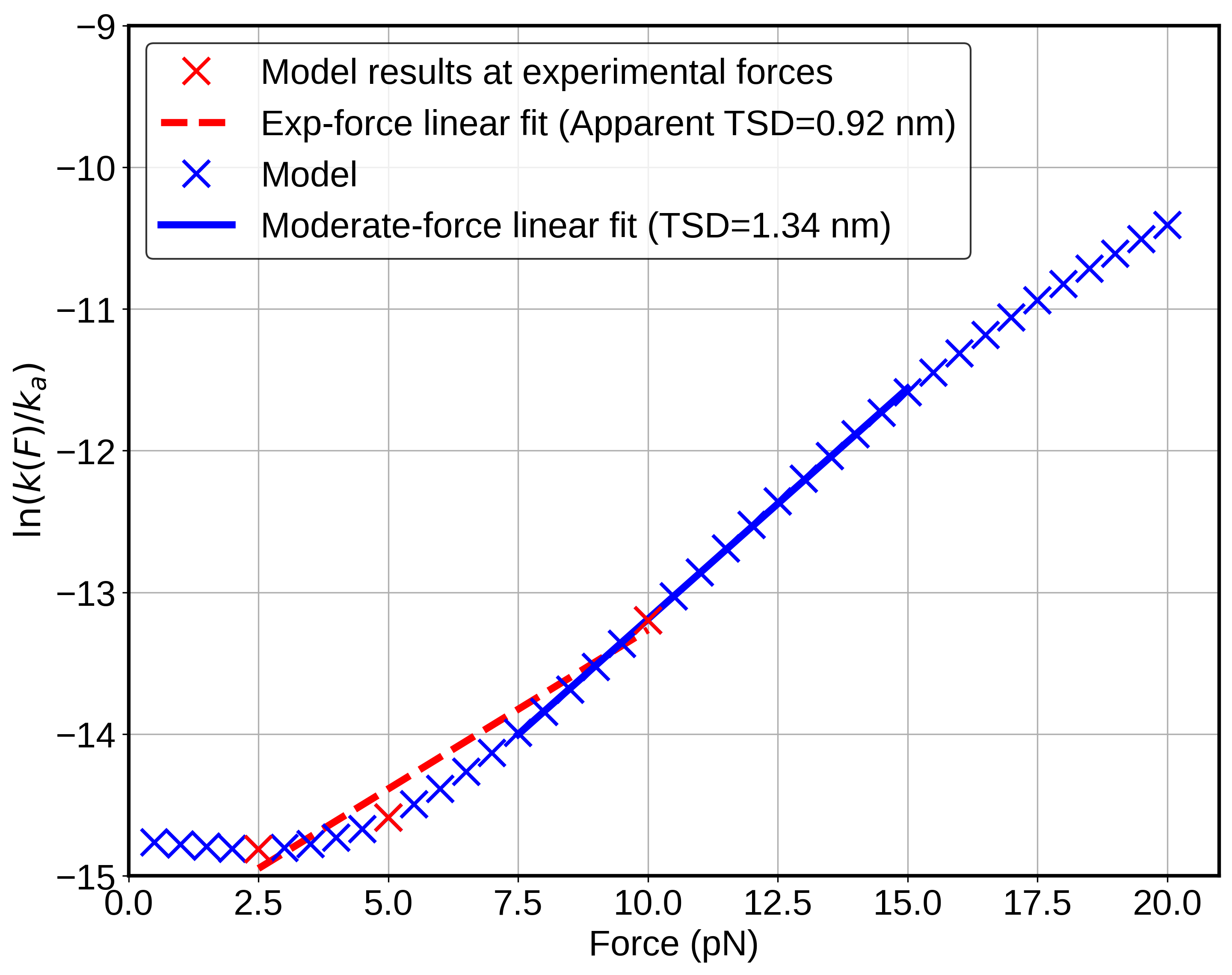}
        \caption{}
        \label{fig:MM-T25-All}
    \end{subfigure}
    \caption{Model results for dark rupture, global temperature of $25^\circ$C. (a) The model result for the three experimental forces $(2.5, 5, 10)$pN with the corresponding Bell's linear fit, and (b) the model result for all forces below $20$pN with a step size of $0.5$pN and the corresponding Bell's linear fit at experimental forces as well as in the moderate-force regime ($7.5-15$pN).}
    \label{fig:MM-T25}
\end{figure}

The system of master equations implementing our model is solved using the helical end to end distance (see \textit{Supporting Information} for detailed calculation) with a $0.34$nm rise, $2$nm diameter, and $10.5$bp/turn for dsDNA~\autocite{Vologodskii_2015} along with $l_{\text{ss}}=0.7$nm as the interphosphate distance and $\lambda_{\text{ss}}=0.77$nm as the persistence length for ssDNA~\autocite{bosco_elastic_2014}. 
The model result for the transition state distance (TSD) using the experimental forces of $2.5$, $5$, and $10$pN is $d_{\text{Dark}}^{\text{Model}}=0.92$nm, which agrees quite well with the experimental TSD at dark rupture ($d_{\text{Dark}}^{\text{Exp-MLE}}=0.85\pm0.20$nm), under Bell's assumption (Fig.~\ref{fig:MM-T25-Three}). \\
In addition to the good agreement obtained for the experimental forces of Kabtiyal et al.~\autocite{kabtiyal_localized_2024}, the model captures two additional important features of the behavior of the reaction off-rate as a function of force.
First, the force dependence was analyzed experimentally by fitting a linear model to $\ln k(F)$, Eq.~(\ref{equ:bell}), using three discrete force values~\autocite{kabtiyal_localized_2024}. However, these three data points (Fig.~\ref{fig:lnkvsf-exp-dark}) show deviations from a linear trend that can be due to either experimental error or intrinsic non-linear behavior in this low-force regime ($\le10$pN).
Evaluating our model at various intermediate forces in this regime supports the latter conclusion as it shows a pronounced curvature (Fig.~\ref{fig:MM-T25-All}) in agreement with experiment, which was also observed in prior studies~\autocite{whitley_elasticity_2017,dna_bow_weak_2023}. These findings do not only validate our model but also point towards careful consideration when applying Bell's formula in the low-force regime as well as interpreting the, generally, \textit{force-dependent} TSD. To this end, we used the term \enquote{apparent transition state distance (Apparent TSD)} in Fig.~\ref{fig:MM-T25}, and we will, from here on, be mostly using it to refer to the distance ($d$) obtained from the slope of the linear fit of the three points of $\ln k(F)$ in this low, non-linear, force regime.\\
Second, since Bell's formula (Eq.~\ref{equ:bell}) assumes strictly linear behavior of $\ln k(F)$, one should only apply it in clearly-linear regimes to extract a meaningful TSD. Evaluating the model results beyond $10$pN shows such linear regime in the moderate-force range ($7.5-15$pN), where one can safely apply Bell's formula (Fig.~\ref{fig:MM-T25-All}). 
This results in a constant TSD per base pair of $1.34\text{nm}/11\text{\text{bp}}\approx0.12$nm/bp, which agrees well with the $0.10$nm/bp found in Strunz et al.~\autocite{strunz_dynamic_1999} using Atomic Force Microscopy (AFM) and the $0.14$nm/bp reported in Whitley et al.~\autocite{whitley_elasticity_2017} using optical tweezers, validating our model in this moderate-force regime. Finally, as expected, the model results in a saturation that starts to emerge at higher forces ($\ge 15$ pN), where the free energy barrier vanishes, rendering the barrier-crossing description underlying Bell's formula itself inapplicable. \\
Overall, the model does capture the observed behavior of the reaction off-rate ($\ln k(F)$) at different force ranges and the resulting TSD with physically grounded polymer models for both dsDNA and ssDNA at those short length scales and buffer conditions.

%%%%%%%%%%%%%%%%%%%%%
\subsection{The dsDNA helical geometry is essential}
\begin{figure}[ht!]
    \centering
    \begin{subfigure}[t]{0.5\textwidth}
        \centering
        \includegraphics[width=\textwidth]{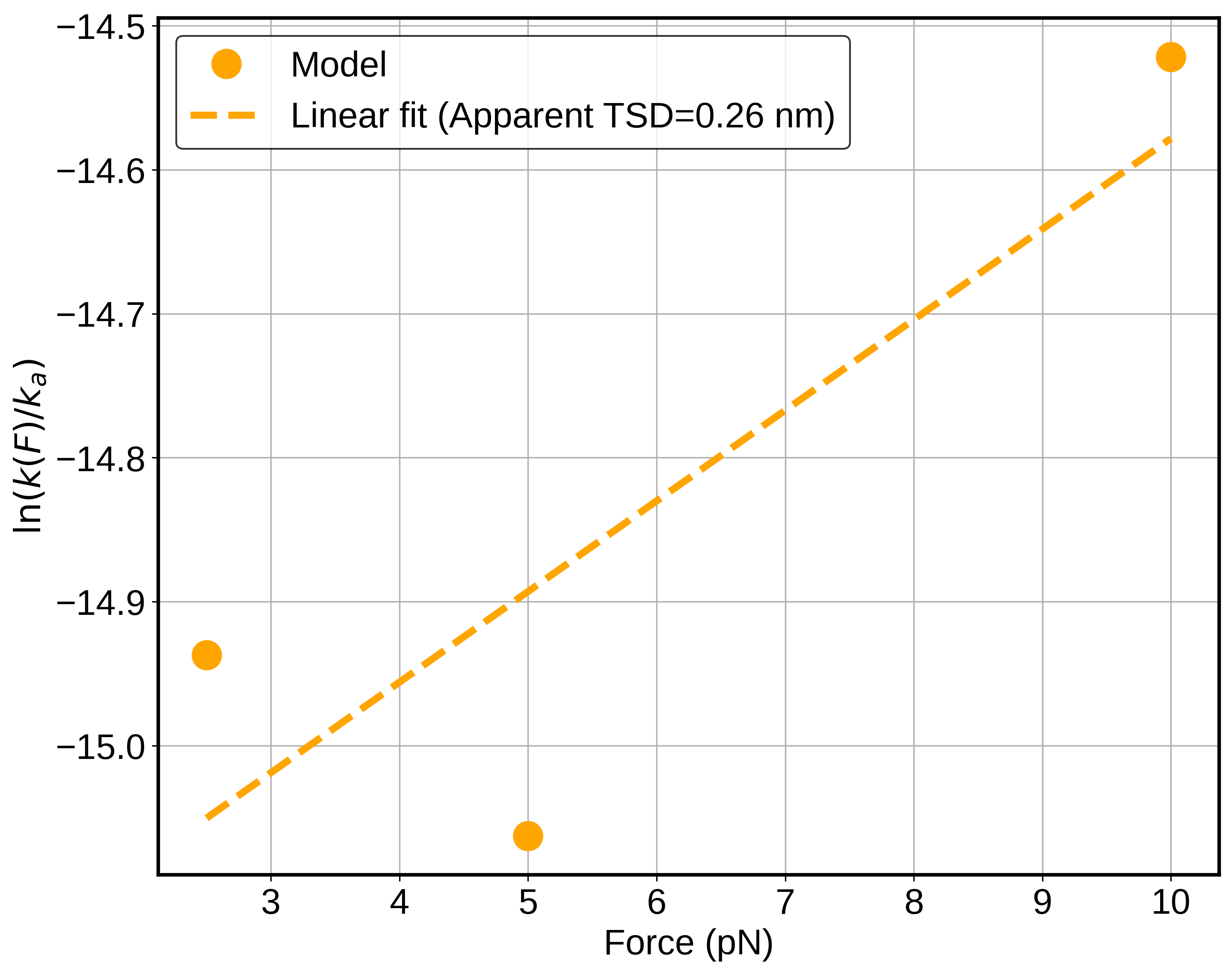}
        \caption{}
        \label{fig:MM-T25-Three-Contour}
    \end{subfigure}%
    ~ 
    \begin{subfigure}[t]{0.5\textwidth}
        \centering
        \includegraphics[width=\textwidth]{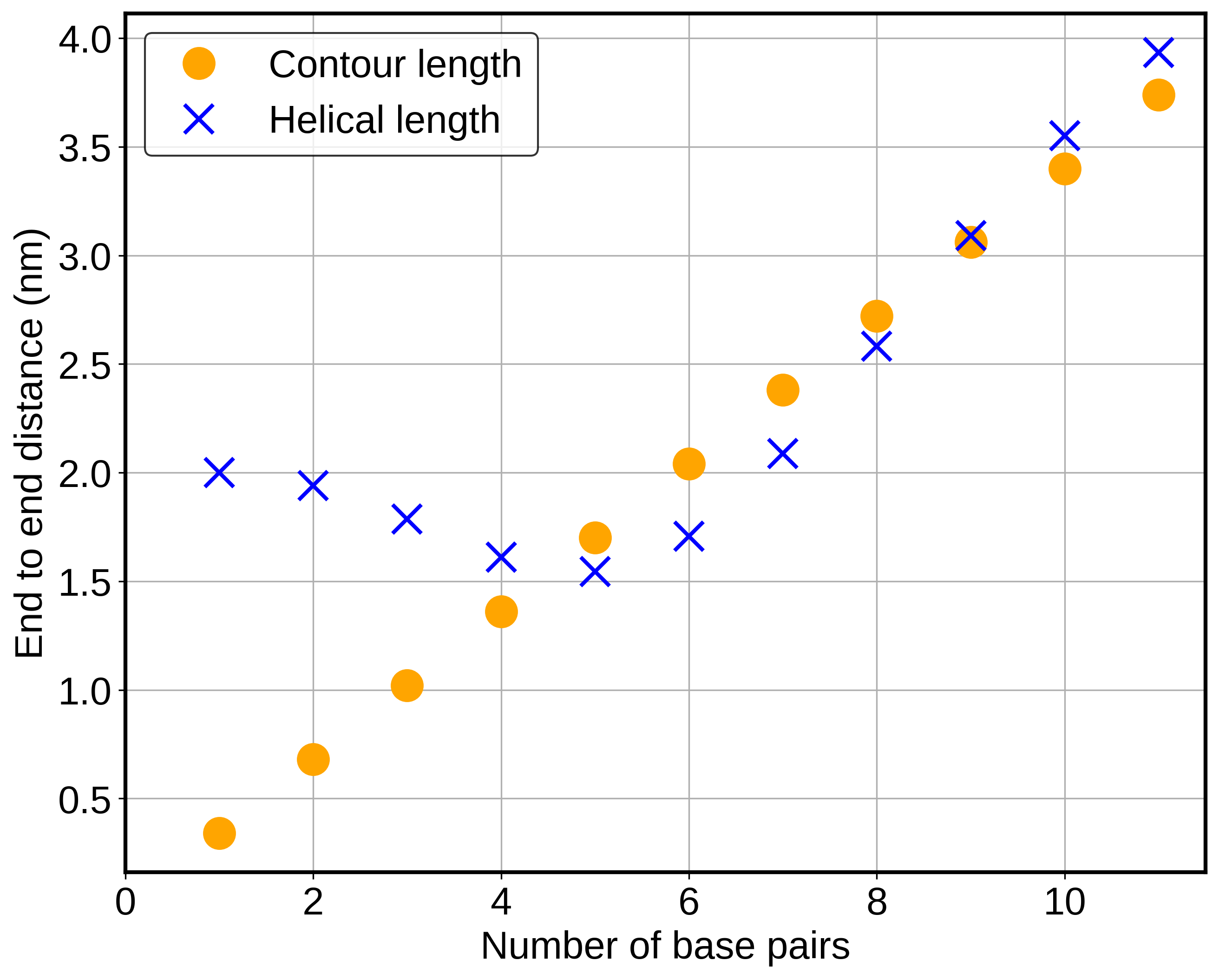}
        \caption{}
        \label{fig:helical-geo}
    \end{subfigure}
    \caption{(a) Model results for dark rupture using contour length as end to end distance of dsDNA, and (b) The duplex end to end distance as a function of the number of base pairs with (Helical length, crosses) and without (Contour length, circles) accounting for the helical geometry. The actual end to end distance with helical geometry is calculated by getting the euclidean distance in 3D space, see \textit{Supporting Information}.}
\end{figure}

When using the rod-like model for dsDNA (Eq.~(\ref{equ:dx_ds}) in the \textit{Methods Section}) the naive choice to use for the end to end distance is the contour length $L_c=0.34\text{nm}\cdot n$, where $n$ is the number of base pairs, which does not account for the three dimensional helical geometry. This results in a huge deviation in the obtained apparent TSD using our model ($d_{\text{Dark}}^{\text{Model-Contour}}=0.26$nm) in comparison to the experimental apparent TSD ($d_{\text{Dark}}^{\text{Exp-MLE}}=0.85\pm0.20$nm), Fig.~\ref{fig:MM-T25-Three-Contour}. This is mainly due to the small length of the considered dsDNA that renders its diameter of $2$nm comparable to its end to end distance ($\le5$nm), Fig.~\ref{fig:helical-geo}. We thus conclude that taking into account the helical geometry of the dsDNA is essential at this short length scale.

%%%%%%%%%%%%%%%%%%%%%
\subsection{The force-extension response of ssDNA}
The polymer description of ssDNA is highly dependent on the fitted force-extension formulas and the buffer conditions. For the former, many experiments~\autocite{rico-pasto_temperature-dependent_2022, bosco_elastic_2014, alemany_determination_2014, whitley_elasticity_2017} fit the Marko-Siggia force-extension formula~\autocite{marko_stretching_1995} to ssDNA while others fit the freely jointed chain~\autocite{smith_overstretching_1996, bosco_elastic_2014}. Depending on the fitted polymer model, the resulting polymer description will be characterized by a different set of parameters, e.g., persistence length vs. Kuhn's length. We will mainly consider experiments fitting the Marko-Siggia force-extension formula and compare this formula to the Global Interpolation Formula (GIF)~\autocite{andersen_stretching_2022} (see \textit{Supporting Information} for exact definition and force-extension comparison) in order to understand the ssDNA polymer description at our length scale. For the latter, the polymer parameters depend on the buffer conditions, particularly, the persistence length~\autocite{bosco_elastic_2014}. We will survey force-extension experiments of similar buffer conditions and/or short length scales to the experimental construct in Kabtiyal et al.~\autocite{kabtiyal_localized_2024} in order to show the robustness of our model with respect to the single-base distance and persistence length variations at our regime of interest.

%%%%%%%%%%%%%%%%%%%%%
\subsubsection{The transition state distance is insensitive to the force-extension behavior of very few nucleotides of ssDNA}
Given the highly flexible nature of ssDNA, it is crucial to know whether the extension of very few nucleotides ($<5$nt) affects the TSD or not. This is important because neither the Marko-Siggia force-extension relation nor the GIF is expected to accurately describe ssDNA in the extreme limit of very few nucleotides, where molecular discreteness and sequence-specific effects become important. 
Indeed, if the TSD depended on these very short extensions, then the use of a more molecularly refined model to describe the force-extension response of those very few nucleotides would be essential. 
Intuitively, the TSD should be more sensitive to the extension at the transition state, which, for the construct in Kabtiyal et al.~\autocite{kabtiyal_localized_2024}, is expected to involve ssDNA with around $10$ nucleotides~\autocite{whitley_elasticity_2017, dna_bow_weak_2023, oxdna_force-induced_2015}. To show that a molecular model is not required to describe the dissociation in Kabtiyal et al.~\autocite{kabtiyal_localized_2024}, we compared our model results under the Marko-Siggia force-extension relation and the GIF. These two models agree near the $10$nt extension while a deviation is observed near the $5$nt extension (see \textit{Supporting Information} for force-extension comparison) so a quantitative disagreement between the two models in our force regime of interest would support the need to use a more molecularly refined model.
To this end, replacing the GIF with the Marko-Siggia force-extension relation in our model yields nearly identical TSD: $0.89$nm at experimental forces (compared to $0.92$nm using the GIF) and $1.34$nm between $7.5$\,pN and $15$\,pN (compared to $1.34$nm using the GIF), respectively (figures not shown). This close agreement indicates that introducing a more molecularly refined description is not expected to either qualitatively alter our conclusions or produce significant quantitative changes within the uncertainty of the present analysis.

%%%%%%%%%%%%%%%%%%%%%
\subsubsection{The model results are robust under different estimates of ssDNA polymer parameters}
\begin{figure}[ht!]
    \centering
    \begin{subfigure}[ht!]{0.5\textwidth}
        \centering
        \includegraphics[width=\textwidth]{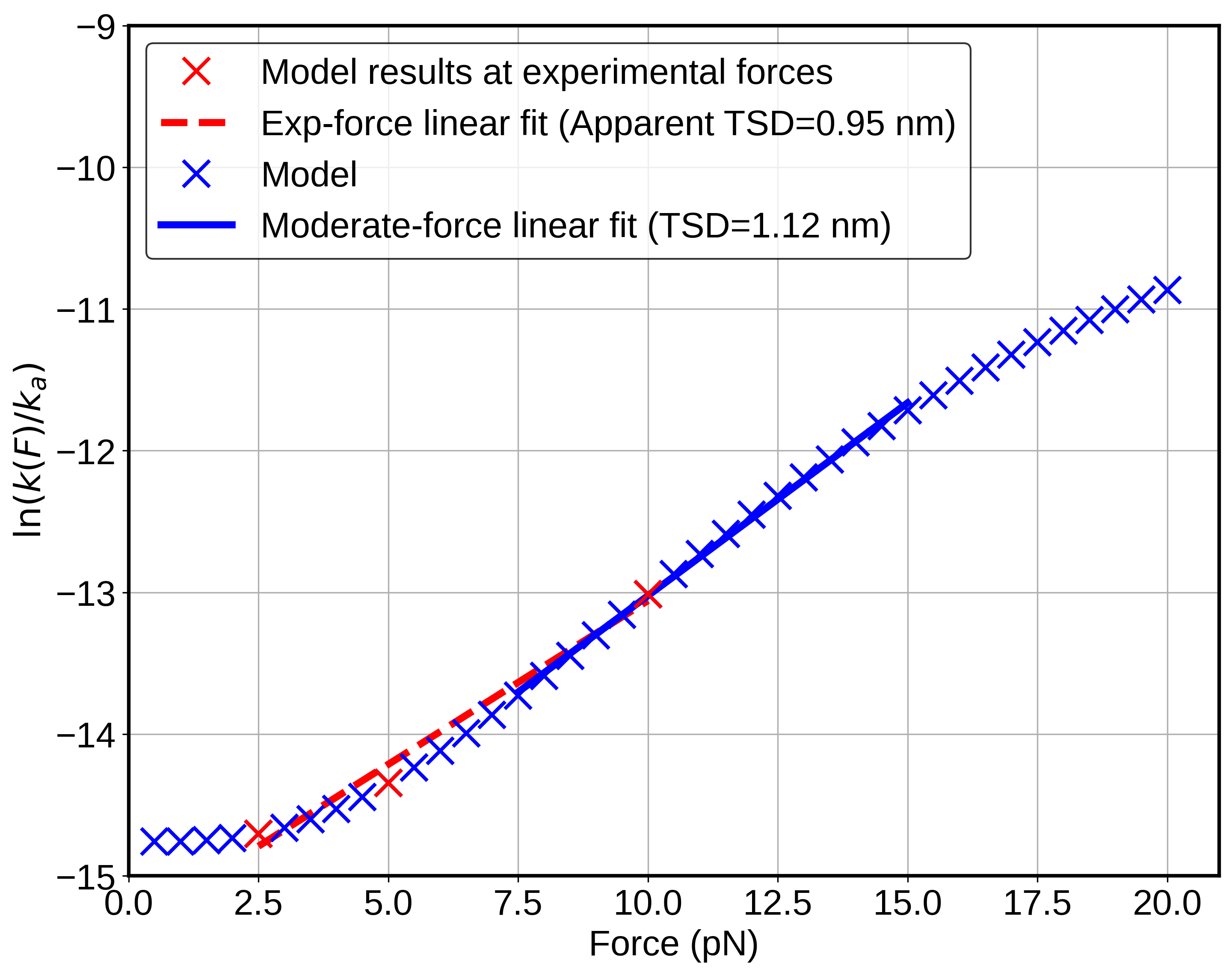}
        \caption{}
    \end{subfigure}%
    ~ 
    \begin{subfigure}[ht!]{0.5\textwidth}
        \centering
        \includegraphics[width=\textwidth]{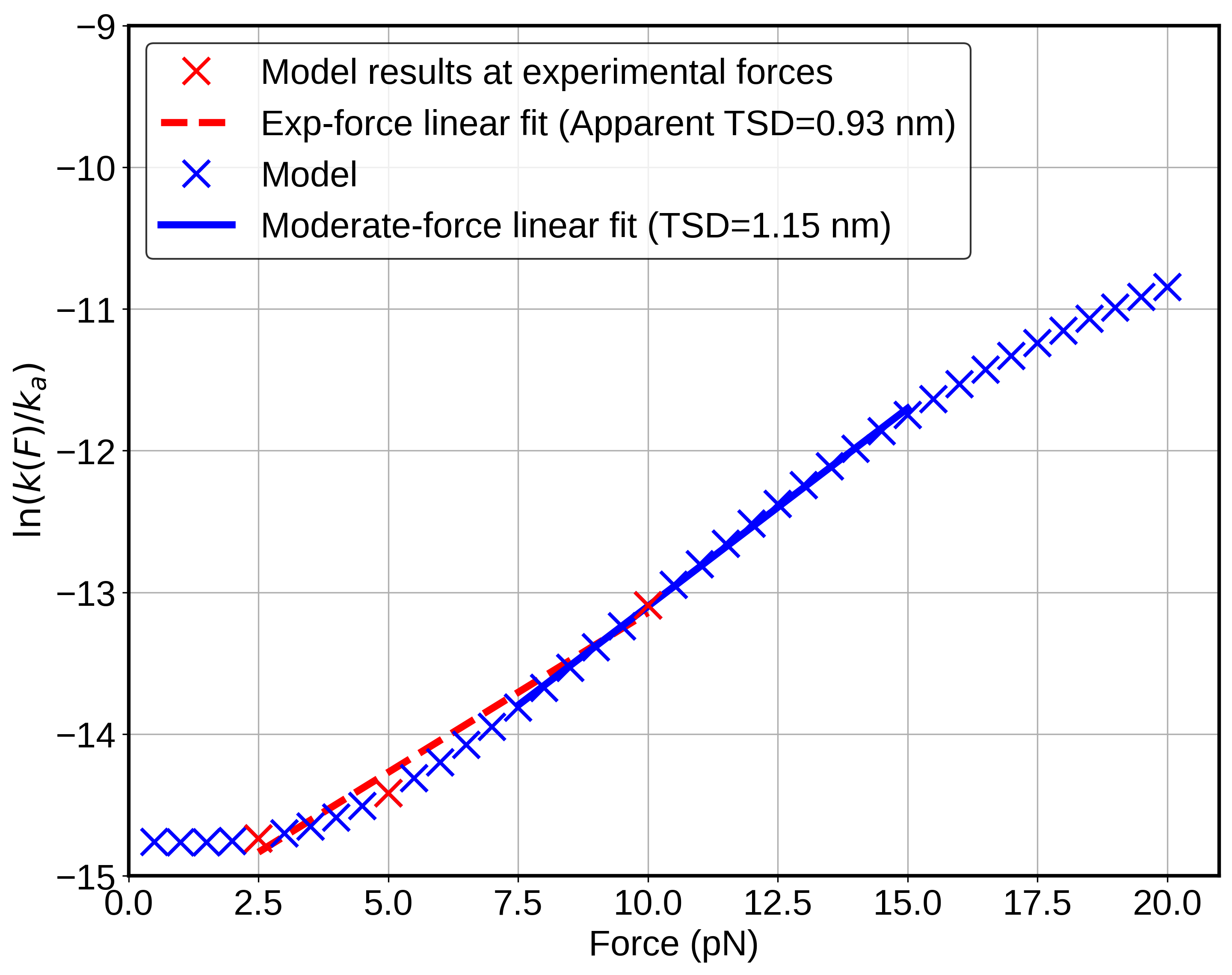}
        \caption{}
    \end{subfigure}
    \caption{The model results for other experimentally-motivated combinations of interphosphate distance and persistence length of ssDNA at experimental salt conditions, $(l_{\text{ss}},\lambda_{\text{ss}})$ respectively: (a) $(0.58, 1.4)$nm, (b) $(0.6, 1.2)$nm. The $(0.7, 0.77)$nm case was shown in Fig.\ref{fig:MM-T25-All}.}
    \label{fig:compare_ss}
\end{figure}

To obtain proper estimates of ssDNA polymer parameters $(l_{\mathrm{\text{ss}}},\lambda_{\text{ss}})$ at the buffer conditions of $5.5$~mM \ce{MgCl2}, $100$~mM \ce{NaCl}, and $0.5\times$TE used in Kabtiyal et al.~\autocite{kabtiyal_localized_2024}, we used prior fits employing the Marko-Siggia force-extension formula from force-extension experiments in similar conditions. 
In all the results presented above, we used estimates of these parameters that were obtained from unzipping assays with long hairpins~\autocite{bosco_elastic_2014} since, to the best of our knowledge, their experiments are the only ones to consistently analyze the behavior of the ssDNA single-base distance and persistence length at different buffer conditions, especially in the presence of magnesium.
In order to obtain the appropriate estimates, we relied on their reported $l_{\mathrm{\text{ss}}} = 0.70 \pm 0.02$~nm and $\lambda_{\text{ss}}$ ranging from $0.75$~nm (at $10$~mM \ce{MgCl2}) to $0.79$~nm (at $4$~mM \ce{MgCl2}), suggesting an intermediate value of $\lambda_{\text{ss}} \approx 0.77$~nm at the buffer conditions in Kabtiyal et al.~\autocite{kabtiyal_localized_2024}   
To evaluate the robustness of our findings with respect to variations in these estimates, we considered two additional force-extension experiments. In particular, we selected one experiment~\autocite{whitley_elasticity_2017} that considered force-shearing of short duplexes, similar to the experimental protocol in Kabtiyal et al.~\autocite{kabtiyal_localized_2024}, and one experiment~\autocite{alemany_determination_2014} that used the same unzipping technique from above but on short hairpins, similar to our length scales. To our knowledge, these are the only experiments to consider force-extension of short ssDNA at our force and salt ranges of interest with Marko-Siggia force-extension fitting.
For the force-shearing experiment~\autocite{whitley_elasticity_2017}, the authors fixed $l_{\mathrm{\text{ss}}} = 0.6$~nm and reported a fitted $\lambda_{\text{ss}} = 1.32 \pm 0.07$~nm at $100$~mM Tris and $100$~mM \ce{NaCl} in the absence of magnesium, while a value of $1.07 \pm 0.05$~nm is obtained in the presence of an additional $20$~mM \ce{MgCl2}. Given the observed relation between the persistence length as a function of salt conditions in this and the previous experiment, an interpolated value of $\lambda_{\text{ss}} \approx 1.2$~nm seems appropriate at the buffer conditions in Kabtiyal et al.~\autocite{kabtiyal_localized_2024} 
The other force-extension experiment~\autocite{alemany_determination_2014} was only conducted at $1$~M \ce{NaCl}. Its authors reported $l_{\mathrm{\text{ss}}} = 0.58 \pm 0.02$~nm and $\lambda_{\text{ss}} = 1.3 \pm 0.2$~nm, which, based on the previous trends, suggests at least $\lambda_{\text{ss}} \approx 1.4$~nm at the buffer conditions in Kabtiyal et al.~\autocite{kabtiyal_localized_2024}
Together, these studies suggest plausible combinations of $(l_{\mathrm{\text{ss}}},\lambda_{\text{ss}})=(0.58,1.4), (0.6,1.2), (0.7, 0.77)\text{nm}$ to describe ssDNA at the buffer conditions of $5.5$~mM \ce{MgCl2}, $100$~mM \ce{NaCl}, and $0.5\times$TE used in Kabtiyal et al.~\autocite{kabtiyal_localized_2024} Remarkably, all three combinations show quite good agreement with the experimentally observed TSDs and with each other (Fig.~\ref{fig:compare_ss}), providing confidence in our estimates and approach. 

%%%%%%%%%%%%%%%%%%%%%
\subsection{The behavior of the transition state distance as a function of temperature}
Analysis of the excited rupture in Kabtiyal et al.~\autocite{kabtiyal_localized_2024} is complicated. The main reason is the distance dependence of the resulting steady-state temperature profile around the AuNP~\autocite{amendola_surface_2017}. This temperature profile yields a different temperature at each base pair location along the duplex, which dynamically changes with each base pair rupture. Also, this resulting temperature at each base pair is highly dependent on the duplex location relative to the AuNP (Fig.~\ref{fig:AuNP}). This relative location is generally random since the two duplexes (DNA-AuNP-DNA) do not have to be along the diameter of the AuNP, and the A11 can bind anywhere on the T23 at each measurement.
Given these complications, we instead first explore our model's response to a global, constant and uniform, increase in temperature, and then we discuss the experimental excited rupture of Kabtiyal et al.~\autocite{kabtiyal_localized_2024}.

%%%%%%%%%%%%%%%%%%%%%
\subsubsection{The model shows negative correlation between temperature and transition state distance}
\begin{figure}[ht!]
    \centering
    \includegraphics[width=0.5\textwidth]{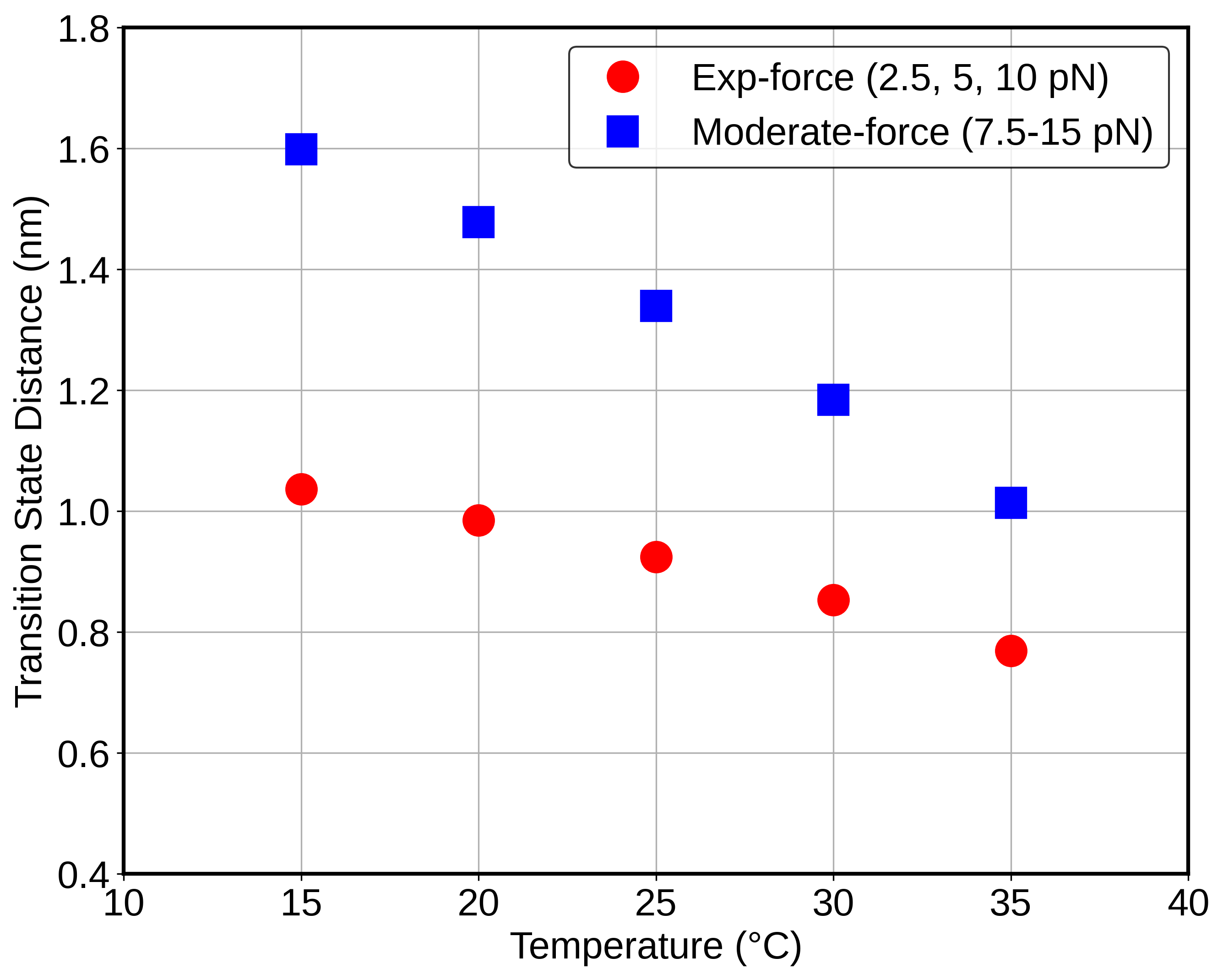}
    \caption{The transition state distance (TSD) obtained from Bell's formula at experimental forces (red dots) and in the moderate-force regime (blue squares) as a function of temperature.}
    \label{fig:compare_T}
\end{figure}

Evaluating our model at different global temperatures shows a decreasing TSD (slope of $\ln k(F)$) as temperature increases in the two ranges of forces we consider (Fig.~\ref{fig:compare_T}). 
This decrease of the TSD with increasing temperature, due to thermal and mechanical destabilization of the duplex, was reported in prior experiments~\autocite{schumakovitch_temperature_2002,kurus_determination_2018}, thus confirming our model's ability to capture global temperature changes and further validating its overall predictive capacity.

%%%%%%%%%%%%%%%%%%%%%
\subsubsection{Careful considerations are needed to  handle the complications of locally-excited rupture}
\begin{figure}[ht!]
    \centering
    \includegraphics[width=0.5\textwidth]{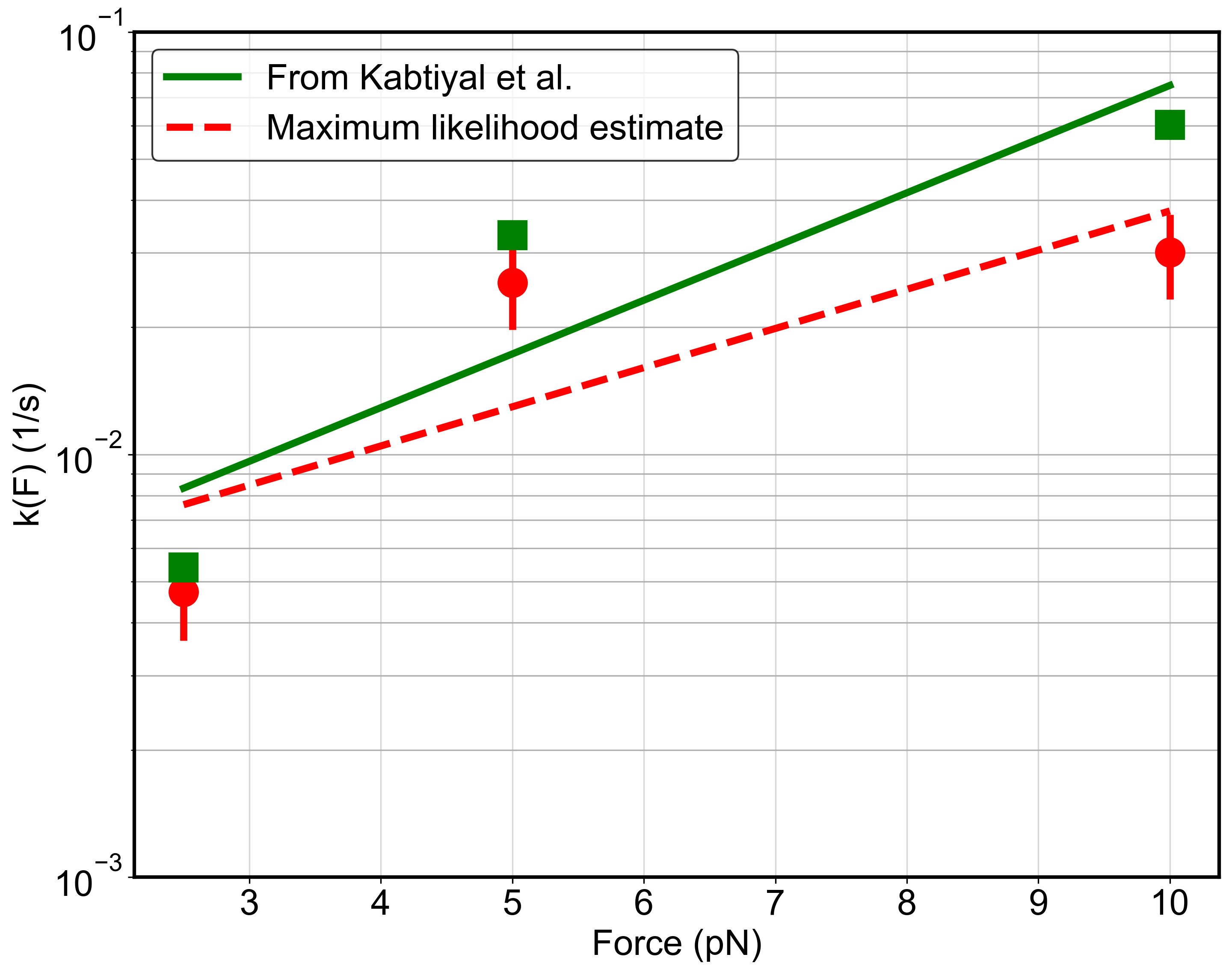}
    \caption{A comparison of the different analyses of the experimental \textit{excited} rupture times, and their corresponding Bell's linear fits (Eq.(\ref{equ:bell})). Circles and dashed line; Maximum-likelihood estimate (MLE), squares and solid line; \textit{excited} rupture results from~\autocite{kabtiyal_localized_2024}}
    \label{fig:lnkvsf-exp-exc}
\end{figure}

At first glance, the excited rupture result in Kabtiyal et al.~\autocite{kabtiyal_localized_2024}, seems to indicate a significant increase in the apparent TSD, $d_{\text{Excited}}^{\text{Exp}}=1.24\pm0.63$nm (green-solid line in Fig.~\ref{fig:lnkvsf-exp-exc}) compared to dark rupture (Fig.~\ref{fig:lnkvsf-exp-dark}), as temperature is increased by localized laser heating. However, in the previous section, we showed that the increase in temperature should lead to a decrease ($\approx0.1$nm per 5$^\circ$C in the low-force range) in the apparent TSD. To tackle this contradiction, we first reanalyzed the excited rupture data using the aforementioned maximum-likelihood analysis. Instead of the sharp increase in the apparent TSD suggested in the original analysis in Kabtiyal et al.~\autocite{kabtiyal_localized_2024}, this reanalysis shows an insignificant change in the apparent TSD between dark ($d^{\text{Exp-MLE}}_{\text{Dark}}=0.85\pm 0.20$nm) and excited ($d^{\text{Exp-MLE}}_{\text{Excited}}=0.90\pm 0.18$nm) ruptures (red-dashed line in Fig.~\ref{fig:lnkvsf-exp-dark} and Fig.~\ref{fig:lnkvsf-exp-exc}). 
\\
Still, the maximum-likelihood analysis does not show the decreasing apparent TSD as temperature increases that we observe using our model. To resolve this, we note that the calculated average excited temperature in Kabtiyal et al.~\autocite{kabtiyal_localized_2024} is around $34^\circ$C. 
At this temperature increase of close to $10^\circ$C, the decrease in the apparent transition distance should just become observable relative to the experimental error of $0.2$nm, according to our model.
The fact that we still do not see an experimentally observable change in the apparent TSD might suggest a lower temperature increase than reported. \\
One indication that the temperature change in the experiment in Kabtiyal et al.~\autocite{kabtiyal_localized_2024} is well below $10^\circ$C is directly inferred from the experimental intercepts in Fig.~\ref{fig:lnkvsf-exp-dark} and Fig.~\ref{fig:lnkvsf-exp-exc}. Those account for temperature and duplex free energies according to Bell's formula (Eq.~(\ref{equ:bell})). Hence, we can use them to obtain a measure of the average temperature at the excited rupture ($T_{\text{Excited}}$)
$$
\Delta\Delta(G_{\text{duplex}}(T)/k_BT)\approx -5.41-(-7.22)=1.81=\frac{1}{k_B}\cdot(\frac{\Delta G_{\text{duplex}}(T_{\text{Dark}})}{T_{\text{Dark}}}-\frac{\Delta G_{\text{duplex}}(T_{\text{Excited}})}{T_{\text{Excited}}}).
$$
Solving this for $T_{\text{Excited}}$ given a dark temperature of $25^\circ$C with the salt-corrected nearest-neighbor parameters~\autocite{santalucia_thermodynamics_2004,owczarzy_predicting_2008} yields an excited temperature of approximately $30^\circ$C. As the model showed (red circles in Fig.~\ref{fig:compare_T}) this excited temperature should cause at most a $0.1$nm drop in the apparent TSD, which is less than the experimental error of $0.2$nm, and thus is not expected to be captured by the experiment.\\
An independent indication of a relatively modest temperature profile induced by localized laser heating follows from a careful analysis of the absorption cross-section, $C_{\mathrm{abs}}$, of the $15~\mathrm{nm}$ AuNP at the excitation wavelength of $561~\mathrm{nm}$~\autocite{jain_calculated_2006}. Estimates based on multiple implementations of Mie scattering theory (see \textit{Supporting Information} for details) consistently yield $C_{\mathrm{abs}}\approx(0.05$--$0.07)\times10^{-15}\,\mathrm{m}^2$, whereas the value reported in Kabtiyal et al.~\autocite{kabtiyal_localized_2024}, $C_{\mathrm{abs}}\approx0.5\times10^{-15}\,\mathrm{m}^2$, is approximately an order of magnitude larger. Using this smaller absorption cross-section leads to a steady-state temperature increase of only $\sim1$--$2^\circ\mathrm{C}$ around the AuNP. \\
Resolving the remaining discrepancy, between the experimentally observed acceleration of rupture kinetics in the excited mode, suggesting about $5^\circ$C difference, and the calculated even more modest $1-2^\circ$C temperature increase due to localized heating, likely requires accounting for additional geometric and photo-thermal complexities inherent to DNA-AuNP-DNA constructs. 
Some of those additional complexities are non-specific laser heating that can result in direct temperature increase of the surrounding medium, and misalignment of the laser spot that may lead to missing the AuNP, both of which are not captured by the simplified photo-thermal description employed here. Based on the observed acceleration of rupture in~\autocite{kabtiyal_localized_2024} with localized laser heating, these additional complications, and the two calculations presented in this section, we believe that the observed temperature rise is sufficient to observe accelerated rupture kinetics, but it is not expected to induce a measurable shift in the apparent TSD beyond the experimental error.

%%%%%%%%%%%%%%%%%%%%%%%%%%%%%%%%%%%%%%%%%%%%%%%%%%%%
\newpage
\section{Conclusions}
We developed a force-dependent kinetic framework for force-induced shear-rupture of short dsDNA and applied it to a DNA-AuNP-DNA construct under constant load. 
Formulating rupture as a master equation problem built from single-base transitions along a force-dependent nucleation-zipper pathway yields a physically grounded description of the dissociation kinetics, and the resulting model reproduces the constant-temperature data while providing a unified interpretation of prior measurements on similarly sheared duplexes across all force regimes.\\
A principal outcome is that the mechanical description of the construct is inseparable from the kinetic analysis. The rod-like model with rod length obtained using the three-dimensional helical geometry of dsDNA is required; replacing the rod length with the typical contour end to end distance produces a significant deviation from experimentally measured transition state distance. For the released ssDNA, different reasonable polymer descriptions yield consistent results in the experimentally relevant regime, though care may still be needed if the construct involves even less base pairs than the one studied here as then the mechanical behavior of the very short ssDNA at the transition state might depend on the molecular details. Together, these observations identify the polymer-mechanical ingredients that must be controlled before assigning molecular meaning to physical parameters inferred from rupture data.\\
Regarding temperature, the framework clarifies the dependence of the inferred kinetic landscape on temperature and, particularly, shows the complications inherent to plasmonically excited rupture. Localized heating, position-dependent temperature profiles along the construct, and geometry-specific factors all require explicit treatment. The present framework provides a pathway for addressing these effects in future studies of AuNP-coupled systems where force and local temperature are tuned simultaneously.\\
Sheared short duplexes are increasingly deployed as force-bearing components in sensors, actuators, and nanoparticle-coupled molecular devices. The kinetic description developed here supports more robust interpretation of existing rupture experiments and more predictive design of force- and temperature-actuated DNA nanostructures, particularly those integrating plasmonic components for on-demand actuation.

%%%%%%%%%%%%%%%%%%%%%%%%%%%%%%%%%%%%%%%%%%%%%%%%%%%%
\newpage
\section{Methods}
%%%%%%%%%%%%%%%%%%%%%%%%%%%%%%%
\subsection{Reaction rates per single-base transition}
\begin{figure}[ht!]
    \centering
    \includegraphics[width=0.5\textwidth]{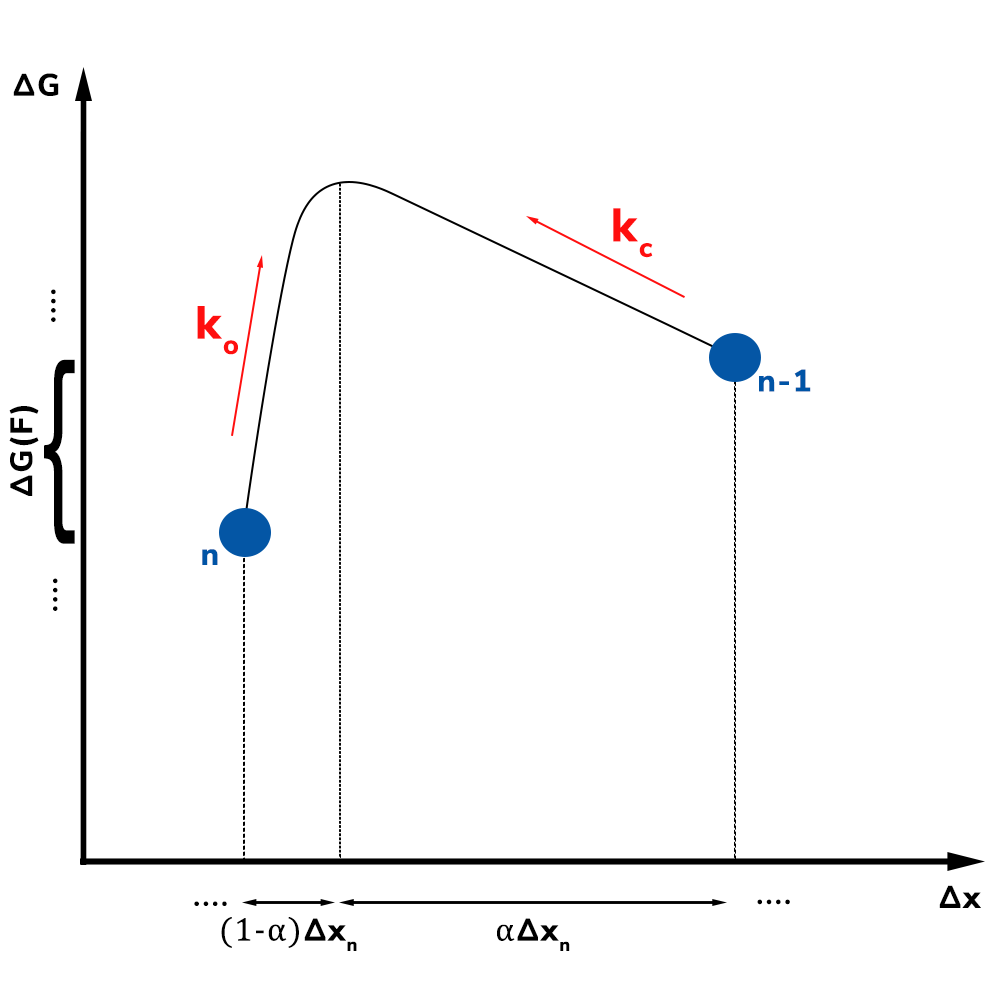}
    \caption{A schematic of a single-base transition barrier between $n$ and $n-1$ closed base pairs. The y-axis is the duplex free energy ($\Delta G$), and the x-axis is the duplex extension ($\Delta x$). $k_c$ and $k_o$ are the single-base closing and opening rates, respectively. $\Delta G(F)$ and $\Delta x_{n}$ are the change in free energy and extension per single-base transition, respectively (Eq.~(\ref{equ:equ-rate})).}
    \label{fig:TS}
\end{figure}

In order to develop the set of equations that govern our kinetic system (Fig.~\ref{fig:model-tree}), one has to specify the transition rates ($k_o,k_c$, and $k_{ot}$) which are dependent on the exact sequence at hand, the mechanical work arising from the distance change due to base pair formation, and the buffer conditions. The ratio of the two rates, $k_c$ and $k_o$, gives the equilibrium constant
\begin{align}
    s_n=k_c/k_o =e^{-\Delta G(F)/k_B T}=e^{-(\Delta G_s-\int_0^F\Delta x_n(f) df)/k_B T}
    \label{equ:equ-rate}
\end{align}
between the $n-1$ base pair duplex and the $n$ base pair duplex~\autocite{vologodskii_dna_2018} with force-dependent free energy~\autocite{dudko_theory_2008,whitley_elasticity_2017}. This comprises multiple components that we will unpack starting from the right-most side of this equation. First, the force-independent free energy ($\Delta G_s(T)=\Delta H_s-T\Delta S_s$) is taken from the nearest-neighbor stacking parameters~\autocite{santalucia_unified_1998,santalucia_thermodynamics_2004}, which are tabulated at 1M monovalent salt concentration, so salt-correction~\autocite{owczarzy_predicting_2008} has to be implemented to match the experimental conditions of Kabtiyal et al.~\autocite{kabtiyal_localized_2024}, see \textit{Supporting Information} for details. The force dependent part of the free energy (the mechanical work or the integral in Eq.~(\ref{equ:equ-rate})) involves the change in the length due to base pair formation
\begin{align}
    \Delta x_n(f)=\Delta x_n^{\text{dsDNA}}(f)+\Delta x_n^{\text{ssDNA}}(f),
    \label{equ:dx}
\end{align}
which, in general, is force-dependent~\autocite{marko_stretching_1995,smith_overstretching_1996}. This change in length is calculated by adding up the individual changes of the ssDNA and the dsDNA portions for the partially ruptured duplex due to the formation of a base pair. \\
We model the ssDNA as a worm like chain~\autocite{marko_stretching_1995,alemany_determination_2014,andersen_stretching_2022} and denote the extension of a strand with $m=N_{\text{bp}}-n$ nucleotides as $\langle z\rangle_m$. Then, the change due to a single base pair formation ($n-1\rightarrow n$) of this worm-like chain is given by
\begin{align}
    \Delta x_n^{\text{\text{ssDNA}}}&=\langle z\rangle_{N_{\text{\text{bp}}}-n}-\langle z\rangle_{N_{\text{\text{bp}}}-(n-1)},
    \label{equ:gif}
\end{align}
where we write it using $n$ (the number of base pairs) instead of $m$ (the number of nucleotides) to have the same notation for both ssDNA and dsDNA. Given the short length scales in the experimental construct ($m\le11$nt), we quantify the force-extension behavior of this worm-like chain through a Global Interpolation Formula (GIF)~\autocite{andersen_stretching_2022} which asymptotically behaves like the well-known Marko-Siggia force-extension formula~\autocite{marko_stretching_1995} for contour lengths $L_m^{\text{ss}}=l_{\text{ss}}\cdot m$ much larger than the persistence length ($L_m^{\text{ss}}\gg \lambda_{\text{ss}}$) but also remains accurate for short lengths ($L_m^{\text{ss}}\gtrsim \lambda_{\text{ss}}$) as is the case for the ssDNA in our system of interest (see the \textit{Supporting Information} for exact definition).\\
For dsDNA, its short length ($n\le11$bp), where the contour length is much shorter than the persistence length ($\approx150$bp)~\autocite{Vologodskii_2015}, suggests a rod-like behavior where the Marko-Siggia force-extension formula does not apply ($L_n^{\text{ds}}\ll \lambda_{\text{ds}}$). Even though the GIF does capture this rod-like regime we chose the rod-like model for dsDNA due to its computational efficiency (see the \textit{Supporting Information} for polymer model comparison).
Hence, the dsDNA portion is modeled as 
\begin{align}
    \Delta x_n^{\text{dsDNA}}(f)=L_n^{\text{ds}}\big[ \coth\big(\frac{f\cdot L_n^{\text{ds}}}{k_BT}\big)-\frac{k_BT}{f\cdot L_n^{\text{ds}}}\big]-L_{n-1}^{\text{ds}}\big[ \coth\big(\frac{f\cdot L_{n-1}^{\text{ds}}}{k_BT}\big)-\frac{k_BT}{f\cdot L_{n-1}^{\text{ds}}}\big],
    \label{equ:dx_ds}
\end{align}
where $L_{n}^{\text{ds}}$ is the end to end distance of a duplex of $n$bp with an interphosphate distance of $l_{\text{ds}}=0.34$nm, where, as shown in the results, the dsDNA end to end distance has to account for its helical geometry~\autocite{Vologodskii_2015} (see \textit{Supporting Information} for the exact calculation).\\
The remaining components of Eq.~(\ref{equ:equ-rate}) are the opening and closing rates per single-base transition ($k_o$,$k_c$), which are key ingredients to the model, Fig.~\ref{fig:model-tree}. In the case where no force is present, it was found that the closing rate is the same for different types of base pairs ($k_c=k_a$, the attempt rate) which, according to Eq.~(\ref{equ:equ-rate}), allows the opening rate to be dependent on the type of the base pair being ruptured ($k_o=k_ae^{\Delta G_s/k_BT}$), i.e., stronger stacks are harder to rupture~\autocite{Vologodskii_2015,santalucia_thermodynamics_2004}. 
In order to define the rates in the presence of force, one has to specify the nature of the transition state between the open and closed base pair states~\autocite{dudko_theory_2008,whitley_elasticity_2017,dna_bow_weak_2023}. In general, given a single-base change of $\Delta x_n$ between the open and closed states, the transition state location can be specified through an adjustable parameter $0\le\alpha\le1$ such that $(1-\alpha)\Delta x_n$ is the distance between the closed state ($n$bp) and the transition state; hence, $\alpha\Delta x_n$ is the distance between the open state ($n-1$bp) and the transition state, Fig.~\ref{fig:TS}. 
This general notion of the distance to the transition state allows us to define force-dependent rates $k_o= k_a e^{\Delta G_s/k_BT}\cdot e^{-(1-\alpha)\int_0^F \Delta x_n(f)df/k_BT}$ and $k_c= k_a e^{\alpha\int_0^F \Delta x_n(f)df/k_BT}$, which maintains the form of the equilibrium constant $s_n$ in Eq.~(\ref{equ:equ-rate}) as required. A physically motivated choice of $\alpha$ is around 1, which means that in order to form a base pair, one has to get the bases in the correct configuration (very close to each other) before they can hybridize, or, looking at it the other way, one has to first break the base pair bonding before the force-extension happens. This implies that the change in extension is primarily affecting the closing transition. Hence, we choose $\alpha=1$ in our computations. We also checked other values of $\alpha$ slightly below 1; these showed an insignificant (<10\%) change in the resultant overall transition state distance (results not shown).\\
In summary, the following form is used for the individual rates in the presence of force
\begin{align}
    k_o&= k_a e^{\Delta G_s/k_BT}\label{equ:ko_rate}\\
    k^{n-1\rightarrow n}_c&= k_a e^{\int_0^F \Delta x_n(f)df}
    \label{equ:kc_rate}
\end{align}
where only the closing rate depends on $n$ \textit{via} $\Delta x_n$.  
Finally, the terminal opening rate is similar to the opening rate in Eq.~(\ref{equ:ko_rate}), but instead involves the non-stacking interactions
\begin{align}
    k_{ot}=k_ae^{(\Delta G_{i}+2\Delta G_{t})/k_BT}
    \label{equ:term-rate}
\end{align}
which includes the terminal AT penalty $\Delta G_{t}$ and the initiation free energy $\Delta G_{i}$, where the factor of $2$ in front of $\Delta G_{t}$ accounts for the terminal AT penalty of the two ends of the 11bp polyA/polyT duplex under investigation~\autocite{santalucia_thermodynamics_2004}, Fig.~\ref{fig:AuNP}. Similar to the stacking interactions, the non-stacking interactions must also be corrected for salt concentration to reflect the experimental conditions accurately (see the \textit{Supporting Information} for detailed salt-correction procedures).

%%%%%%%%%%%
\subsection{Master equations approach for the time evolution of the probabilities of different states}
\label{sec:MEs}
Given the kinetic scheme developed in prior sections, the time evolution is given \textit{via} the master equations for the probability of each state in this system~\autocite{master_cosmic_1940,hanggi_reaction-rate_1990,vankampen2007spp}. The time evolution of these probabilities can be compacted into two main formulae. For the ruptured (ssDNA) state, we have
\begin{gather}
    \frac{dP_{\text{ssDNA}}(t)}{dt}=k_{ot}\sum_{(i,j)|i+j=N_{\text{bp}}-1}P_{(i,j)}(t)
    \label{equ:transitions1}
\end{gather}
where $P_{(i,j)}(t)$ is the probability of the state with $i$ open bases from the left side and $j$ open bases from the right side at time $t$, and $k_{ot}$ is the terminal rate. For all other states we obtain
\begin{gather}
    \frac{dP_{(i,j)}}{dt}=\
    \underbrace{k_o\!\left[\mathbf{1}\{i>0\}P_{(i-1,j)}(t)+\mathbf{1}\{j>0\}P_{(i,j-1)}(t)\right]}_{\text{incoming by opening}}
    +\underbrace{\mathbf{1}\{n>1\}\,k_c^{\,n-1\to n}\!\left[P_{(i+1,j)}(t)+P_{(i,j+1)}(t)\right]}_{\text{incoming by closing}}\nonumber\\[2mm]
    -\underbrace{\big(2k_o\,\mathbf{1}\{n>1\}+k_{ot}\,\mathbf{1}\{n=1\}\big)P_{(i,j)}(t)}_{\text{outgoing by opening}}
    -\underbrace{\big(\mathbf{1}\{i>0\}+\mathbf{1}\{j>0\}\big)\,k_c^{\,n\to n+1}\,P_{(i,j)}(t)}_{\text{outgoing by closing}}.
    \label{equ:transitions2}
\end{gather} 
where $\mathbf{1}\{\cdot\}$ is an indicator function (1 if true, 0 otherwise) and $n=N_{\text{bp}}-i-j$ is the number of closed base pairs in the state $(i,j)$. A more extended way of writing this system of master equations without the indicator function is shown in the \textit{Supporting Information}.\\
This linear system of equations can then be mapped into a matrix differential equation
\begin{gather}
    \frac{d\textbf{P}(t)}{dt}=\textbf{T}\cdot\textbf{P}
    %\leftrightarrow\frac{dP_{(i,j)}(t)}{dt}=\sum_{(m,n)}(\mathbf{T})_{((i,j),(m,n))}P_{(m,n)}(t)
    \label{equ:pt}
\end{gather}
where \textbf{T} is the transition matrix with elements from Eqs. (\ref{equ:transitions1}) and (\ref{equ:transitions2}), and $\textbf{P$(t)$}$ is the vector of time-dependent probabilities (All simulation codes and data are publicly available\autocite{hussein_2026_20597758}). The initial condition is taken as the full duplex state, $P_{(0,0)}=1$ and $P_{(i,j)}=0\ \forall (i,j)\ne(0,0)$. This matrix differential equation can then be solved using standard differential equations techniques, where the solution yields the probability of each state at a given moment in time~\autocite{strang2006linear}. In particular, we select a thousand equally spaced time points between 0 and $t_e$, where $t_e$ is the thermal dissociation time (see \textit{Supporting Information} for exact definition) and use them to obtain the probability of the ruptured state, $P_{\text{ssDNA}}(t)$. 
Given the extensive number of data points from the numerical solution, we fit a single-exponential to the probability of the ruptured state ($P_{\text{ssDNA}}(t)\approx1-e^{-k(F)t}$), which yields the reaction off-rate ($k(F)$) that we can then use to obtain the transition state distance (e.g., through Bell's equation, Eq.~(\ref{equ:bell})) and compare to the experimental results as shown earlier.\\
An important consideration in modeling the experimental system in Kabtiyal et al.~\autocite{kabtiyal_localized_2024} is that the AuNP construct, Fig.~\ref{fig:AuNP}, contains two DNA duplexes connected in series between the central particle and the side beads. From an experimental perspective, rupture of either duplex leads to complete detachment of the construct, meaning that the entire system is considered ruptured once one side fails. 
To incorporate this in our model, we first calculate the probability of a single duplex being in the ruptured state at time $t$, $P_{\text{ssDNA}}(t)$. The probability that a single duplex has \textit{not} ruptured by time $t$ is then $1 - P_{\text{ssDNA}}(t)$. Because the two duplexes behave independently, the probability that both remain intact is $(1 - P_{\text{ssDNA}}(t))^2$; hence, the probability that at least one duplex has ruptured by time $t$ is given by
\begin{equation}
    P_c(t)=1 - (1 - P_{\text{ssDNA}}(t))^2 \approx 1 - e^{-2kt}.
\end{equation}
This corresponds to an effective rate constant twice that of a single duplex, or effectively halving the characteristic rupture time as expected for two independent rupture pathways. In terms of the standard $\ln k$ vs. $F$ form (Eq.~(\ref{equ:bell})), this doubles the observed off-rate, resulting in a shift of the $\ln k(F)$ intercept by $\ln 2$, while the transition state distance remains unaffected.

\section{Acknowledgments}
This research was partially supported by the Center for Emergent Materials, an NSF MRSEC, under award number DMR-2011876. GitHub Copilot was used to enhance the source code readability, and Claude was used in writing the initial draft of the introduction section.

\section*{Supporting Information}

Supporting Information Available: Maximum-likelihood-estimate reanalysis of experimental 
rupture times; global interpolation formula (GIF) and comparison to other polymer models; 
calculation of the end-to-end distance of dsDNA with helical geometry; estimation of 
the absorption cross-section for 15 nm AuNP at 561 nm wavelength; extended system 
of master equations; salt correction of nearest-neighbor parameters; and mathematical 
methods for solving the system of master equations.\\

%%%%%%%%%%%%%%%%%%%%%%%%%%%%%%%%%%%%%%%%%%%%%%
%%%%%%%%%%%%%%%%%%%%%%%%%%%%%%%%%%%%%%%%%%%%%%
\newpage
\begin{center}
    \Large Supporting Information for \enquote{A kinetic model of shear-induced rupture of short dsDNA}\\
    \large Ayman Hussein and Ralf Bundschuh
\end{center}

\section*{Maximum-likelihood analysis of experimental data with right-censoring}
Here, we briefly describe the maximum-likelihood estimate method (MLE) to analyze experimental rupture times and obtain reaction off-rates. For right-censored data (observations start at $t=0$ and end at the latest at $t=t_c=600s$ for the experiment of Kabtiyal et al.~\autocite{kabtiyal_localized_2024}), the observed likelihood function is~\autocite{lawless_statistical_2003}
\begin{align*}
    L=\prod_{i=1}^{n}f(t_i)^{\delta_i}S(t_i)^{1-\delta_i}
\end{align*}
where $t_1,...,t_n$ are the censored rupture times, $f(t)$ is the probability density function of the rupture time $t$, $S(t)=\Pr(t^\prime\geq t)=\int_{t}^{\infty}f(t^\prime)dt^\prime$ is the survivor function, and $\delta_i=1$ if the instance ruptured before the cut-off time ($t_i<t_c$) while $\delta_i=0$ if no rupture was observed in this window ($t_i>t_c$).
In the case of exponentially distributed rupture times, the probability density is given by
\begin{align*}
    f(t)=ke^{-k t}
\end{align*}
where $k$ is the rate constant or, alternatively, $\tau=1/k$ is the mean lifetime of the molecular construct. This yields a survivor function of $$S(t)=e^{-kt}.$$
The likelihood function for the experiment of Kabtiyal et al.~\autocite{kabtiyal_localized_2024} is thus
\begin{align*}
    L(k)=k^me^{-k\sum_{i=1}^{n}t_i}
\end{align*}
where $n=20$ is the total number of instances and $m\le n$ is the number of ruptured instances ($\sum_{i=1}^n\delta_i$). Maximizing this likelihood function with respect to $k$ yields the maximum-likelihood estimate
\begin{align}
    \hat{k}=\frac{1}{\hat{\tau}}=\frac{m}{\sum_{j=1}^{m}t_j+(n-m)\cdot t_c}
    \label{equ:k_hat}
\end{align}
for the rate constant (or the inverse of the life time).\\
The uncertainty in the logarithm of the estimated rate constant can be derived by noting that 
$\ln(\hat{k}) = -\ln(\hat{\tau})$, where $\hat{\tau}$ denotes the estimated mean lifetime. 
Because $\hat{\tau}$ is proportional to the sum of $m$ independent exponentially distributed rupture times (the denominator in Eq.~(\ref{equ:k_hat})), and given that the sum of exponentially distributed random variables follows a Gamma distribution~\autocite{lawless_statistical_2003}, it follows that $\hat{\tau}$ itself is Gamma distributed, 
$\hat{\tau} \sim \Gamma(m, k)$. 
The variance of $\ln(\hat{\tau})$, and thus of $\ln(\hat{k})$, is then obtained directly from the variance of the Log--Gamma distribution, which is given by the trigamma function, $\psi^{(1)}(m)$.

%%%%%%%%%%%%%%%%%%%%%%%%%%%%%%%%%%%%%%%%%%%%%%%%%%%%%%%%%%%%%%%%%%
\section*{Finite-Length dsDNA and ssDNA Polymer Models under Force}
\begin{figure}[ht!]
    \centering
    \includegraphics[width=0.7\linewidth]{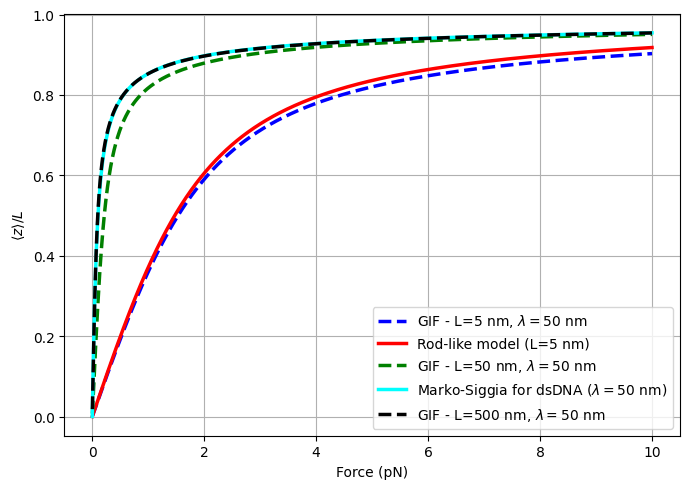}
    \caption{Comparison of the global interpolation formula (GIF), the rod-like model, and the Marko-Siggia interpolation formula. The dashed curves (from right to left) are the GIF with increasing end to end distance. The rightmost solid curve is for the rod-like model while the leftmost one is for the Marko-Siggia force-extension formula.}
    \label{fig:WLC}
\end{figure}

Here, we explain our choice of (i) the rod-like model (Eq.~(5) in the main text) for describing dsDNA force-extension behavior, and (ii) the global interpolation formula (GIF), Eq.~(\ref{equ:global-interpolation}) below, for ssDNA rather than the widely adopted Marko-Siggia formula~\autocite{marko_stretching_1995} for both. \\
On the one hand, we expect the dsDNA to be nearly straight over its entire contour length ($L_{ds}$) of $\le11$bp since this is far shorter than the typical persistence length ($L_{ds}\ll \lambda_{ds}$) of dsDNA ($\approx$50nm or $\approx$150bp), while the Marko-Siggia formula has been derived in the limit of molecules much larger than their persistence length~\autocite{Vologodskii_2015}. 
On the other hand, the ssDNA is quite flexible and has a persistence length ($\lambda_{ss}$) on the order of its single-base distance ($L_{ss}\gtrsim \lambda_{ss}$ for the short sequences considered here). Accounting for the different $L/\lambda$ can be captured using a single global interpolation formula~\autocite{andersen_stretching_2022}.
The GIF approximates the exact Worm-Like-Chain (WLC) force-extension behavior within 1\% accuracy over the full range of force and length scales. It is given by
\begin{align}
    \frac{\langle z\rangle}{L}
    &= \coth\!\left(f^{\dagger}\right) - \frac{1}{f^{\dagger}}\nonumber\\
    \text{where}\quad f^{\dagger}
    &= \tilde f\,
    \frac{1 + A_{1}(\frac{\lambda}{L})\tilde f + A_{2}(\frac{\lambda}{L})\tilde f^{2} + A_{3}(\frac{\lambda}{L})\tilde f^{3}}
    {1 + a_{0}\tilde f + b_{0}\tilde f^{2} + A_{3}(\frac{\lambda}{L})D(\zeta)\langle\textbf{R}_{\mathrm{KP}}^{2}\rangle\tilde f^{3}/L^{2}},
    \label{equ:global-interpolation}
\end{align}
$\langle z\rangle$ is the extension of the DNA, $L$ is the contour length, $\lambda$ is the persistence length, $\langle\textbf{R}_{\mathrm{KP}}^{2}\rangle=2\lambda LK(\frac{\lambda}{L})$ with $K(x)=1-x-xe^{-\frac{1}{x}}$, and $\tilde f=\frac{\langle\textbf{R}_{\mathrm{KP}}^{2}\rangle f}{k_BTL}$ is the reduced force, where $f$ is the force. The remaining functions are 
\begin{align*}
    A_{1}(x) &= \frac{a_{0} + a_{1}K(x)}{1 + a_{1}'K(x)} ,& &A_{2}(x) = \frac{b_{0} + b_{1}K(x)}{1 + b_{1}'K(x)} \\
    A_{3}(x) &= \frac{c_{0} + c_{1}K(x)}{1 + c_{1}'K(x)} ,\text{and}&  &D(\zeta)=[1+\zeta\coth(\zeta)]/2
\end{align*}
where $\zeta=L\sqrt{\frac{f}{\lambda k_BT}}$ and the lowercase coefficients are given in Table 1 of Andersen et al.~\autocite{andersen_stretching_2022}\\
In Fig.~\ref{fig:WLC} we show the normalized extension $\left(\langle z\rangle/L\right)$) from the GIF in the (i) flexible limit ($L/\lambda>1$) corresponding to ssDNA and long dsDNA, (ii) stiff limit ($L/\lambda<1$) corresponding to short dsDNA. We also show the rod-like model ($f^{\dagger}=\frac{fL}{k_BT}$ in Eq.~(\ref{equ:global-interpolation})) and the Marko-Siggia force-extension formula for comparison.\\ 
The rod-like model agrees well with the GIF in the stiff limit, while the Marko-Siggia model (which is independent of contour length) performs better in the flexible limit but fails in the stiff regime relevant to short dsDNA. Given its accuracy in this regime and greater computational efficiency, we use the rod-like model for dsDNA force-extension. For ssDNA, we use the GIF instead, as it is both simpler to integrate than the Marko-Siggia formula and
more accurate when $L\gtrsim\lambda$, which corresponds to a few nucleotides of ssDNA.

%%%%%%%%%%%%%%%%%%%%%%%%%%%%%%%%%%%%%%%%%%%%%%%%%%%%%%%%%%%%%%%%%%
\section*{dsDNA end to end distance with helical geometry}
\begin{figure}[ht!]
    \centering
    \begin{subfigure}[t]{0.49\textwidth}
        \centering
        \includegraphics[height=3in]{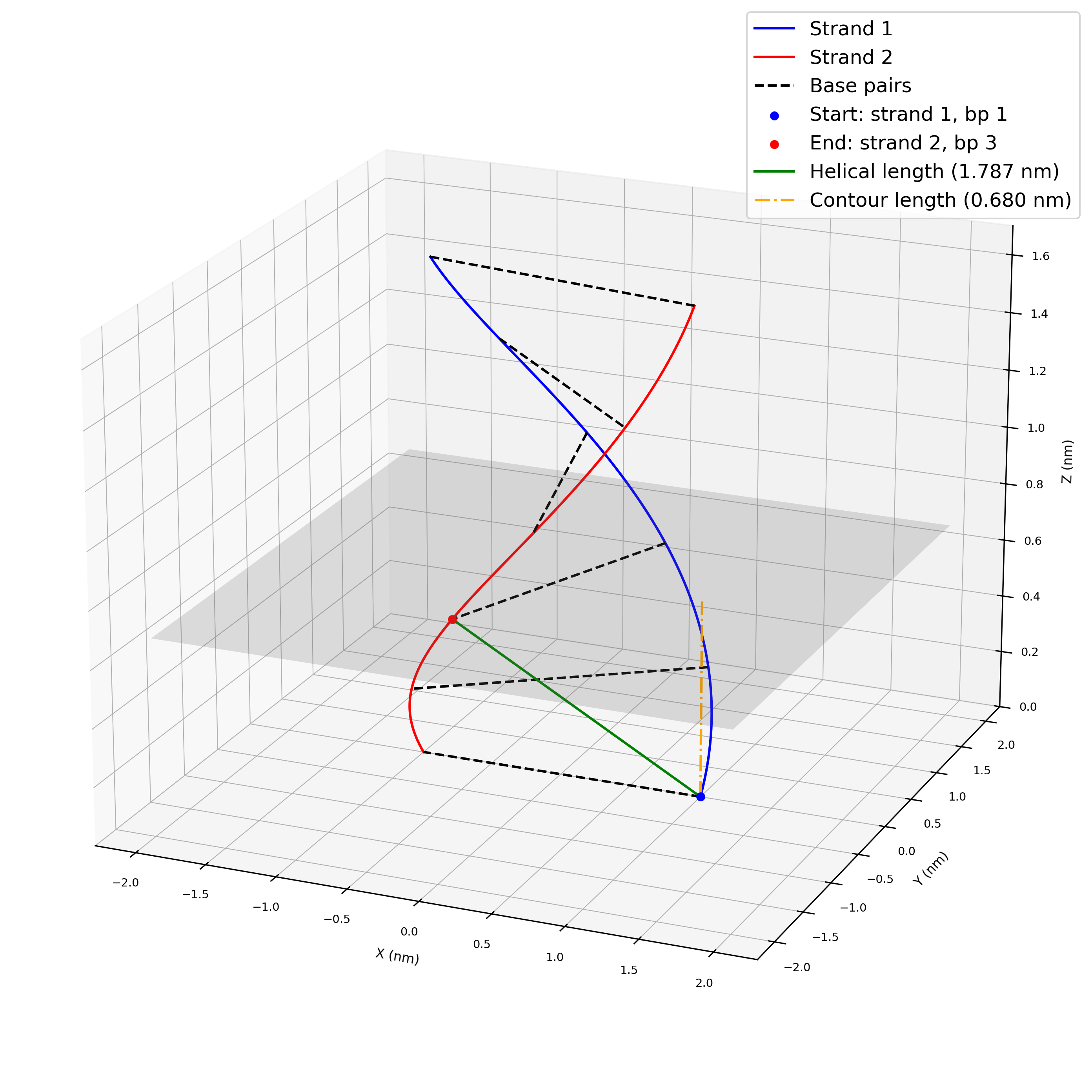}
    \end{subfigure}%
    ~ 
    \begin{subfigure}[t]{0.49\textwidth}
        \centering
        \includegraphics[height=3in]{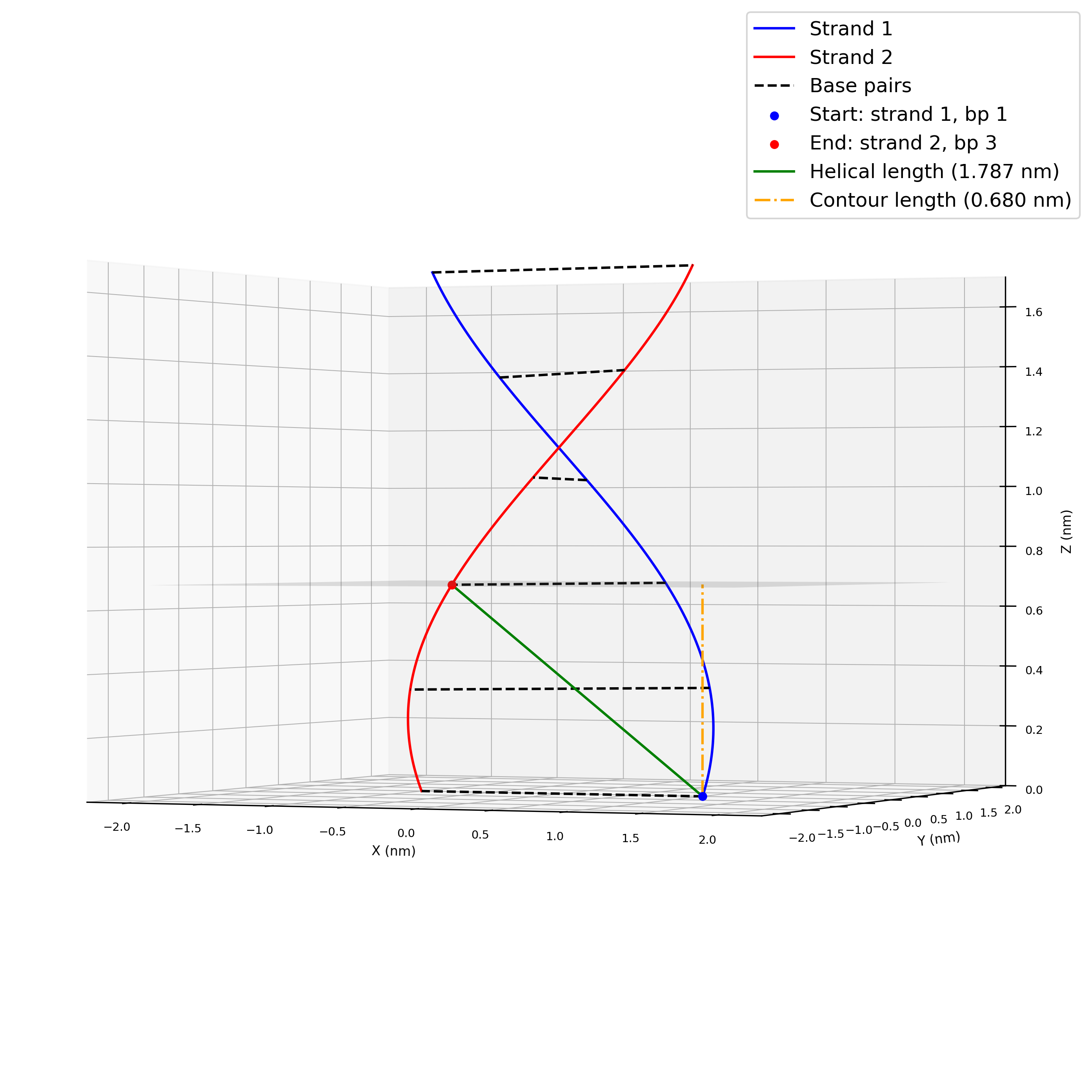}
    \end{subfigure}
    \caption{Two 3D views of the helical geometry of dsDNA. The starting point (blue) is on the first base of the first helical backbone (blue) and the second point (red) is on the third base of the second helical backbone (red). The green-solid line shows the helical length while the orange dot-dashed line shows the contour length between these two bases.}
    \label{fig:helixvscontour}
\end{figure}

The helical end to end distances used in Figs.~(4), (6), and (7) were calculated by explicitly accounting for the three-dimensional B-DNA helical geometry with standard parameters: $2$~nm diameter, $0.34$~nm rise per base pair, and a helical pitch of $10.5$~bp per turn~\autocite{Vologodskii_2015}. The two backbones are parameterized as helices
\begin{equation}
\vec{r}_1(\phi) = \begin{pmatrix} R\cos(\phi) \\ R\sin(\phi) \\ \frac{p}{2\pi}\phi \end{pmatrix}\quad \text{and} \quad
\vec{r}_2(\phi) = \begin{pmatrix} -R\cos(\phi) \\ -R\sin(\phi) \\ \frac{p}{2\pi}\phi \end{pmatrix},
\end{equation}
where $R = 1.0$~nm is the helix radius, $p =0.34\text{nm}\times10.5= 3.57$~nm is the helical pitch, and $\phi$ is the parametric variable representing the angular position along the helix.
For a given base number $n$ on each strand, the parameter value is calculated as
\begin{equation*}
\phi(n) = \frac{2\pi}{p} \cdot \frac{p}{10.5} \cdot n.
\end{equation*}
The position of base $n_1$ on the first strand and base $n_2$ on the second strand are given by $\vec{r}_1(\phi(n_1))$ and $\vec{r}_2(\phi(n_2))$, respectively. The end to end distance between these points is given by the norm of their vector difference
\begin{equation}
d_{e2e} = \left|\vec{r}_2(\phi(n_2)) - \vec{r}_1(\phi(n_1))\right|,
\end{equation}
while the contour length, representing the vertical distance along the helical axis, is
\begin{equation}
L_c = \left|\frac{p}{2\pi}\phi(n_2) - \frac{p}{2\pi}\phi(n_1)\right| = 0.34|n_2 - n_1| \text{ nm}.
\end{equation}
Fig.~\ref{fig:helixvscontour} shows the 3D comparison of the two end-to-end distances, and Fig.~5b shows the exact values for the end-to-end distance along the $11$bp duplex. 

%%%%%%%%%%%%%%%%%%%%%%%%%%%%%%%%%%%%%%%%%%%%%%%%%%%%%%%%%%%%%%%%%%
\section*{Absorption cross-section estimates for the $15$~nm AuNP at $561$~nm wavelength}
The optical absorption and scattering behavior of gold nanospheres were examined to calculate the size-dependent trends governing their photothermal response. Within the framework of Mie theory, the extinction, absorption, and scattering cross-sections are related to the particle diameter through well-defined scaling laws in the dipole regime~\autocite{jain_calculated_2006}. Here, we first use these relations to extrapolate the cross-section data reported in Jain et al.~\autocite{jain_calculated_2006} to estimate the absorption cross-section of the $15\mathrm{nm}$ AuNP employed in the experiment of Kabtiyal et al.~\autocite{kabtiyal_localized_2024} at 561nm wavelength ($C_{\text{abs}}^{15\text{nm}}$). Then, we report direct calculations of $C_{\text{abs}}^{15\text{nm}}$ using some of the publicly available implementations of Mie theory.

\subsection*{An upper limit estimate of $C_{\text{abs}}^{15\text{nm}}$ based on the size dependency of gold nanospheres cross-section}
The absorption cross-section of $20$, $40$, and $80$~nm gold nanospheres was obtained at different wavelengths~\autocite{jain_calculated_2006} using Mie scattering theory calculations. The main conclusions found were (i) the extinction cross-section (sum of absorption and scattering cross sections) increases with the nanosphere volume, which is a consequence of the dipole mode being dominant when applying Mie scattering theory at this scale, and (ii) the scattering to absorption ratio increases with the nanosphere size due to the increase in radiative damping, where the ratio ($C_{\text{sca}}/C_{\text{abs}}$) is about $0.01$ for the $20$~nm gold nanosphere. 
They also reported a volume normalized absorption cross-section ($C_{\text{abs}}/V$) for the $20$~nm nanosphere of about $\mu_{abs}=73.72\mu$m$^{-1}$ at resonance wavelength ($521$~nm).  Given the size dependency of the extinction cross-section, the scattering to absorption ratio, and the size normalized absorption cross-section of the $20$~nm gold nanosphere, we can obtain an estimate of the absorption cross-section of the $15$~nm gold nanosphere used in~\autocite{kabtiyal_localized_2024} at $521$~nm wavelength, which is around its resonance wavelength due to the weak dependency of the resonance wavelength on the diameter of the nanospheres~\autocite{jain_calculated_2006}. \\
First, the scattering to absorption ratio of the $15$~nm AuNP will be lower than the value of $0.01$ for the $20$~nm AuNP given the reported size dependency, so we can assume the approximate relation of $C_{\text{ext}}=C_{\text{abs}}+C_{\text{sca}}\approx C_{\text{abs}}$ for both particles. Hence, at $521$~nm wavelength, we can estimate $C_{\text{abs}}^{15\text{nm}}$ using the volume dependency as:
\begin{align}
    C_{\text{abs}}^{15\text{nm}}\approx\frac{V^{15\text{nm}}}{V^{20\text{nm}}}C_{\text{abs}}^{20\text{nm}}=1.3\times10^{-16}\text{m}^2
    \label{equ:cabs-upper}
\end{align}
where $C_{\text{abs}}^{20\text{nm}}=\mu_{\text{abs}}\cdot\frac{4}{3}\pi (10\times10^{-9}\text{m})^3=3.1\times10^{-16}\text{m}^2$ ($>C_{\text{abs}}^{15\text{nm}}$) is the absorption cross-section of the $20$~nm AuNP at $521$~nm~\autocite{jain_calculated_2006}. This serves as an upper limit on the absorption cross-section of the $15$~nm AuNP near resonance wavelength, which is larger than what it should be at off-resonance wavelengths~\autocite{jain_calculated_2006}, particularly at the $561$nm excitation used in Kabtiyal et al.~\autocite{kabtiyal_localized_2024}.

\subsection*{Direct estimate of the absorption cross-section of the $15$~nm AuNP at $561$~nm wavelength}
Several independent implementations of Mie theory were used to directly estimate the absorption cross-section, $C_{\text{abs}}^{15\text{nm}}$, of a $15~\mathrm{nm}$ gold nanosphere at an excitation wavelength of $561~\mathrm{nm}$. Calculations based on standard online Mie solvers, including the NanoComposix Mie calculator~\autocite{nanocomposix} and the OMLC Mie code~\autocite{omlc_mie} combined with tabulated bulk gold refractive index data~\autocite{refractiveindex}, yield values in the range $C_{\mathrm{abs}}^{15\,\mathrm{nm}}\simeq(0.5$--$0.7)\times10^{-16}\,\mathrm{m}^2$. Additional calculation using MiePlot software~\autocite{mieplot} further confirms that the absorption cross-section of a $15~\mathrm{nm}$ AuNP at $561~\mathrm{nm}$ lies near $0.6\times10^{-16}\,\mathrm{m}^2$. Hence, the various estimates are consistent with each other, with the reported size-dependent trends~\autocite{jain_calculated_2006}, and with the upper bound derived in Eq.~(\ref{equ:cabs-upper}).

%%%%%%%%%%%%%%%%%%%%%%%%%%%%%%%%%%%%%%%%%%%%%%%%%%%%%%%%%%
\section*{Extended system of master equations}
As mentioned in the main text, one can categorize the different master equations based on the number and types of allowed transitions from a given state, resulting in 6 main categories. These are summarized here with the associated examples from the 4bp rupture case, Fig.~3 in main text.\\
The ssDNA state can neither transition into other states through opening nor through closing which we denote by \{0C,0O\}, where O means opening and C means closing. Transitions into the ssDNA state come from breaking the last existing base pair from any of the single base pair states. It thus evolves according to
\begin{gather*}
    \frac{dP_{\text{ssDNA}}(t)}{dt}=k_{ot}\sum_{(i,j)|i+j=N_{bp}-1}P_{(i,j)}(t)
\end{gather*}
where $P_{(i,j)}(t)$ is the probability of the state with $i$ open bases from the left side and $j$ open bases from the right side at time $t$, $N_{bp}$ is the number of base pairs in the initial duplex, and $k_{ot}$ is the terminal rate as defined in the main text. Hereafter, we omit the explicit time dependence $(t)$ in the probabilities for clarity, with the understanding that each probability remains a function of time. \\
The initial duplex state has \{0C,2O\} outgoing transitions and thus evolves according to
\begin{gather*}
    \frac{dP_{(0,0)}}{dt}=k_c^{N_{bp}-1\rightarrow N_{bp}}\left[P_{(1,0)}+P_{(0,1)}\right]-2k_oP_{(0,0)}
\end{gather*}
where $k_c$ and $k_o$ are the closing and opening rates as defined in the main text (Eqs.~(6) and (7)), and the superscript on $k_c$ indicates that the closing event occurs from $N_{bp}-1$ to $N_{bp}$ closed base pairs.\\
The edge terminal states (the last base pair to break is on the edges of the initial duplex (e.g., (0,3) and (3,0) in Fig.~3 of the main text) have \{1C,1O\} outgoing transitions, so they evolve according to
\begin{gather*}
    \frac{dP_{(0,N_{bp}-1)}}{dt}=k_oP_{(0,N_{bp}-2)}-(k_c^{1\rightarrow2}+k_{ot})P_{(0,N_{bp}-1)}\\
    \frac{dP_{(N_{bp}-1,0)}}{dt}=k_oP_{(N_{bp}-2,0)}-(k_c^{1\rightarrow2}+k_{ot})P_{(N_{bp}-1,0)}
\end{gather*}
The internal terminal states (the last base pair to break is not on either edge of the initial duplex (e.g., (1,2) and (2,1) in Fig.~3 of the main text) have \{2C,1O\} outgoing transitions. They correspond to $(i,j)$ that satisfy $i+j=N_{bp}-1$ with $i\ge1$ and $j\ge1$, and thus evolve according to
\begin{gather*}
    \frac{dP_{(i,j)}}{dt}=k_o\left[P_{(i-1,j)}+P_{(i,j-1)}\right]-(k_{ot}+2k_c^{1\rightarrow2})P_{(i,j)}.
\end{gather*}
The edge non-terminal states (they have the edge base pair from the initial duplex (e.g., (0,1), (1,0), (0,2), and (2,0) in Fig.~3 of the main text) have \{1C,2O\} outgoing transitions and evolve according to
\begin{gather*}
    \frac{dP_{(0,j)}}{dt}=k_c^{N_{bp}-(j+1)\rightarrow N_{bp}-j}\left[P_{(0,j+1)}+P_{(1,j)}\right]+k_oP_{(0,j-1)}\\
    -(2k_o+k_c^{N_{bp}-j\rightarrow N_{bp}-(j-1)})P_{(0,j)}\\
    \frac{dP_{(i,0)}}{dt}=k_c^{N_{bp}-(i+1)\rightarrow N_{bp}-i}\left[P_{(i+1,0)}+P_{(i,1)}\right]+k_oP_{(i-1,0)}\\
    -(2k_o+k_c^{N_{bp}-i\rightarrow N_{bp}-(i-1)})P_{(i,0)}
\end{gather*}
where $i$ and $j$ are greater than or equal 1 and strictly less than $N_{bp}-1$. \\
The remaining non-terminal states have \{2C,2O\} outgoing transitions (e.g., (1,1) in Fig.~3 of the main text) and evolve according to
\begin{gather*}
    \frac{dP_{(i,j)}}{dt}=k_o\left[P_{(i-1,j)}+P_{(i,j-1)}\right]+k_c^{N_{bp}-(i+j+1)\rightarrow N_{bp}-(i+j)}\left[P_{(i+1,j)}+P_{(i,j+1)}\right]\\
    -2(k_o+k_c^{N_{bp}-(i+j)\rightarrow N_{bp}-(i+j-1)})P_{(i,j)}.
\end{gather*}

%%%%%%%%%%%%%%%%%%%%%%%%%%%%%%%%%%%%%%%%%%%%%%%%%%%%%%%%%%
\section*{Salt correction of stacking and non-stacking Entropies}
Base pairing free energies are typically measured under standard conditions of $1$~M monovalent salt (\ce{Na}$^+$) and no divalent ions~\autocite{santalucia_thermodynamics_2004}. In contrast, the experiments described in Kabtiyal et al.~\autocite{kabtiyal_localized_2024} were performed in $0.5\times$TE buffer containing $100$~mM NaCl, $5.5$~mM \ce{MgCl2}, and $0.05$\% Tween $20$. Because both monovalent and divalent cations stabilize DNA duplexes, the difference in ionic conditions necessitates applying corrections to the measured thermodynamic parameters, based on established DNA melting analysis models~\autocite{owczarzy_predicting_2008}.\\
The presence of $0.5\times$TE results in having $5$~mM of Tris$^+$ that gets added to the overall monovalent cations concentration. Also, $0.5$~mM EDTA reduces the free Magnesium concentration by $0.5$~mM, so the total concentrations of monovalent cations and free Magnesium become 
\begin{align}
    \text{[Mon$^+$]}&=\text{[Tris$^+$]+[Na$^+$]}=105\text{mM}=0.105\text{M}\label{equ:mono_conc},\\
    \text{and}\quad\text{[Mg$^{2+}$]}&=[\text{Mg$^{2+}$}]_{Total}-[\text{EDTA}]=5.5\text{mM}-0.5\text{mM}=5\text{mM}=0.005\text{M}\label{equ:div_conc}.
\end{align}
Here, we follow the salt correction approach from Owczarzy et al.~\autocite{owczarzy_predicting_2008} where various short ($<30$bp) DNA duplexes were melted at different concentrations of monovalent and divalent cations. According to the various conditions studied, one has to first decide the regime of dominance of monovalent and/or divalent cations. This is captured by the parameter $R=\sqrt{\text{[Mg$^{2+}$]}}/\text{[Mon$^+$]}$ which requires accounting for the total monovalent cation concentration [Mon$^+$] and the free Magnesium concentration  [Mg$^{2+}$]. 
For the concentrations in Eqs.~(\ref{equ:mono_conc}) and (\ref{equ:div_conc}) this results in  $R=\frac{\sqrt{0.005}}{0.105}\approx0.673$, which indicates that the experiment is in the regime where both divalent and monovalent cations are important. This requires the use of a divalent-dominant correction for the total entropy of hybridization with monovalent-corrected coefficients, namely
\begin{align}
     \Delta S^{\circ}\left(\text{Salt}\right)=&\Delta S^{\circ}\left(1 \mathrm{M} \ \mathrm{Na}^{+}\right)\nonumber\\
     &+\Delta H^{\circ}\left\{a+b \ln \left[\mathrm{Mg}^{2+}\right] +\frac{e+f \ln \left[\mathrm{Mg}^{2+}\right]+g\left(\ln \left[\mathrm{Mg}^{2+}\right]\right)^2}{2\left(N_{\mathrm{bp}}-1\right)}\right\}
    \label{equ:ds_mg}
\end{align}
where we focus on polyA/polyT sequences so that the G/C content is set to zero in the original formula (Eq.~
$(22)$ in Owczarzy et al.~\autocite{owczarzy_predicting_2008}). Here, $\Delta H^{\circ}$ and $\Delta S^{\circ}\left(1 \mathrm{M} \ \mathrm{Na}^{+}\right)$ are the standard enthalpy and entropy of the full duplex hybridization at $1$~M Na$^+$~\autocite{santalucia_thermodynamics_2004}, $\Delta S^{\circ}\left(\text{Salt}\right)$ is the salt corrected entropy of the full duplex hybridization at the given cation concentration, $N_{bp}$ is the number of base pairs in the duplex, and the coefficients (in units of K$^{-1}$) are
\begin{align*}
    a&=3.92\times10^{-5}(0.843-0.352\sqrt{\text{[Mon$^+$]}}\times\ln\text{[Mon$^+$]})\\
    b&=-9.11\times10^{-6},\ e=-4.82\times10^{-4},\ f=5.25\times10^{-4},\ \text{and}\\
    g&=8.31\times10^{-5}(0.486-0.258\ln\text{[Mon$^+$]}+5.25\times10^{-3}(\ln\text{[Mon$^+$]})^3)
\end{align*}
where some of them ($a$ and $g$) depend on the monovalent concentration \text{[Mon$^+$]} in M.\\
Since the model relies on the nearest-neighbor stacking parameters $\Delta H_s$ and $\Delta S_s$, and non-stacking parameters $\Delta H_{\text{Terminal AT penalty}}$, $\Delta H_{\text{Initiation}}$, $\Delta S_{\text{Terminal AT penalty}}$, and $\Delta S_{\text{Initiation}}$, we aim to extract corrections for these individual parameters from Eq.~(\ref{equ:ds_mg}). \\
To this end, we write the duplex total hybridization entropy and enthalpy as
\begin{align*}
    \Delta S^{\circ}(\text{Salt})&=(N_{bp}-1)\cdot\Delta S^{\circ}_s(\text{Salt})+\Delta S^{\circ}_{\text{non}}(\text{Salt}),\ \text{and}\\
    \Delta H^{\circ}&=(N_{bp}-1)\cdot\Delta H^{\circ}_s+\Delta H^{\circ}_{\text{non}}
\end{align*}
where, since polyA/polyT sequences are considered, $\Delta S^{\circ}_s=\Delta S^{\circ}_{\text{AA/TT}}$ for all ($N_{bp}-1$) stacks and similarly for $\Delta H^{\circ}_s=\Delta H^{\circ}_{\text{AA/TT}}$, $\Delta H^{\circ}$, where the enthalpy is taken to be independent of salt~\autocite{santalucia_thermodynamics_2004}, and for simplicity we combine the non-stacking contributions into a single term for entropy $\Delta S^{\circ}_{\text{non}}=2\Delta S^{\circ}_{\text{Terminal AT penalty}}+\Delta S^{\circ}_{\text{Initiation}}$ and for enthaply $\Delta H^{\circ}_{\text{non}}=2\Delta H^{\circ}_{\text{Terminal AT penalty}}+\Delta H^{\circ}_{\text{Initiation}}$~\autocite{santalucia_thermodynamics_2004}. Substituting these into Eq.~(\ref{equ:ds_mg}) yields
\begin{align}
    (N_{bp}-1)\cdot\Delta S^{\circ}_s(\text{Salt})\nonumber\\
    +\Delta S^{\circ}_{\text{non}}(\text{Salt})=
    &(N_{bp}-1)\cdot\{\Delta S^{\circ}_s(\text{1M Na$^+$})+\Delta H^{\circ}_s\cdot\left(a+b\ln[\text{Mg$^{2+}$}]\right)\}\nonumber\\
    & +\big\{\Delta S^{\circ}_{\text{non}}(\text{1M Na$^+$})+\Delta H^{\circ}_s\cdot\left(\frac{e+f\ln[\text{Mg$^{2+}$}]+g(\ln[\text{Mg$^{2+}$}])^2}{2}\right)+\nonumber\\
    &~~~~~~~\Delta H^{\circ}_{\text{non}}\cdot\left(a+b\ln[\text{Mg$^{2+}$}]\right)\big\}\nonumber\\
    &+\mathcal{O}\left(\frac{1}{N_{bp}-1}\right).\label{equ:ds_full}
\end{align}
To extract salt correction for the internal stacking contribution, one has to take the infinite duplex limit to cancel end effects. This can be obtained by dividing Eq.~(\ref{equ:ds_full}) by ($N_{bp}-1$) and then taking the limit of $N_{bp}\rightarrow\infty$, which yields
\begin{align}
    \Delta S^{\circ}_s(\text{Salt})&=\Delta S^{\circ}_s(\text{1M Na$^+$})+\Delta H^{\circ}_s\cdot\left(a+b\ln[\text{Mg$^{2+}$}]\right).
    \label{equ:ds_correct_near}
\end{align}
Then the non-stacking contribution is given by the remaining part
\begin{align}
    \Delta S^{\circ}_{\text{non}}(\text{Salt})\approx&\Delta S^{\circ}_{\text{non}}(\text{1M Na$^+$})+\Delta H^{\circ}_{\text{non}}\cdot\left(a+b\ln[\text{Mg$^{2+}$}]\right)\nonumber\\
    &+\Delta H^{\circ}_s\cdot\left(\frac{e+f\ln[\text{Mg$^{2+}$}]+g(\ln[\text{Mg$^{2+}$}])^2}{2}\right)
    \label{equ:ds_correct_non}.
\end{align}
Thus, applying those corrections with the concentrations used in Kabtiyal et al.~\autocite{kabtiyal_localized_2024} Eqs.~(\ref{equ:mono_conc}) and (\ref{equ:div_conc}) for the stacking and non-stacking entropies yields
\begin{align}
    \Delta S^{\circ}_s(\text{Salt})&=-0.0920\ \text{kJ/mol.K}\\
    \Delta S^{\circ}_{\text{non}}(\text{Salt})&=0.0502\ \text{kJ/mol.K}
    \label{equ:ds_correct}
\end{align}
which are the values used to obtain the free energies in the opening rates, $k_o$ and $k_{ot}$ (Eqs.~(7) and (8) in the main text).

%%%%%%%%%%%%%%%%%%%%%%%%%%%%%%%%%%%%%%%%%%%%%%%%%%%%%%%%%%%%%%%%%%
\section*{Solving the system of master equations}
In this section, we describe our approach to solving the system of master equations in order to obtain the time-dependent probability of the ruptured state, \( P_{\mathrm{ssDNA}}(t) \). As discussed in the main text, the system can be expressed as the matrix differential equation  
\begin{gather}
    \frac{d\mathbf{P}(t)}{dt} = \mathbf{T} \cdot \mathbf{P}(t),
    \label{equ:poft}
\end{gather}
where \(\mathbf{T}\) is an \( N \times N \) transition matrix containing the transition rates among $N$ different states (from Eq.~(10) in the main text), and $\mathbf{P}(t)$ is the \( N \times 1 \) vector of probabilities.  
Assuming the system initially occupies the fully bound duplex state \((0,0)\), the initial probability vector is  
\begin{align}
    \mathbf{P}_0 =
    \begin{bmatrix}
        1\\
        0\\
        \vdots\\
        0
    \end{bmatrix},
    \label{equ:p0vec}
\end{align}
which is an \( N \times 1 \) column vector. The formal solution to Eq.~(\ref{equ:poft}) is then given by  
\begin{align}
    \mathbf{P}(t) = e^{\mathbf{T}t} \, \mathbf{P}_0,
\end{align}
where \( e^{\mathbf{T}t} \) denotes the matrix exponential of \( \mathbf{T}t \)~\autocite{strang2006linear}.\\
One way to evaluate this is by diagonalizing $\mathbf{T}$ using standard linear algebra methods~\autocite{strang2006linear}, where $\mathbf{T}$ can be rewritten using its eigenvalues and eigenvectors
$\mathbf{T}=\mathbf{U}\mathbf{D}\mathbf{U}^{-1}$ 
such that 
$
\mathbf{U}=
\begin{bmatrix}
       \vec{v}_1  & \vec{v}_2 & \cdots & \vec{v}_N
\end{bmatrix}
$ 
is the matrix with $N\times1$ eigenvectors ($\vec{v}_i$) as columns, $\mathbf{U}^{-1}$ is its inverse ($\mathbf{U}\mathbf{U}^{-1}=\mathbf{I}_{N\times N}$), and $\mathbf{D}$ is a diagonal matrix with the diagonal elements as its eigenvalues ($\lambda_i$)
$$
\mathbf{D}=
\begin{bmatrix}
        \lambda_1 & 0 &  \cdots & \cdots & 0\\
        0 & \lambda_2 &    & & \vdots\\
        \vdots &  &  \ddots  & & \vdots\\
        \vdots &  & &   \lambda_{N-1} & 0 \\
         0  & \cdots & \cdots & 0 & \lambda_N
\end{bmatrix}.
$$ 
The advantage of diagonalization is how it simplifies the matrix exponential $e^{\mathbf{T}t}$ since
\begin{align*}
    e^{\mathbf{T}t} = e^{\mathbf{U}\mathbf{D}\mathbf{U}^{-1}t} 
    & = \mathbf{U}e^{\mathbf{D}t} \mathbf{U}^{-1} = \mathbf{U}\begin{bmatrix}
        e^{\lambda_1t} & 0 &  \cdots & \cdots & 0\\
        0 & e^{\lambda_2t} &    & & \vdots\\
        \vdots &  &  \ddots  & & \vdots\\
        \vdots &  & &   e^{\lambda_{N-1}t} & 0 \\
         0  & \cdots & \cdots & 0 & e^{\lambda_Nt}
\end{bmatrix}\mathbf{U}^{-1}.
\end{align*}
For the specified initial condition in Eq.~(\ref{equ:p0vec}), the probability of the ruptured state at time \( t \), corresponding to the \( N^{\text{th}} \) component of \( \mathbf{P}(t) \), is then given by  
\begin{align}
    P_{\mathrm{ssDNA}}(t) = [\mathbf{U}e^{\mathbf{D}t} \mathbf{U}^{-1}]_{(N,1)}=\sum_{m=1}^{N}(\vec{v}_N)_me^{\lambda_mt}(\vec{v}^\prime_m)_1
    \label{equ:p_ss}
\end{align}
where $(\vec{v}_N)_m$ is the $m^{th}$ component of the $N^{th}$ eigenvector (i.e., the last column in $\mathbf{U}$), and $(\vec{v}^\prime_m)_1$ is the first component of the $m^{th}$ column of $\mathbf{U}^{-1}$ (i.e., the first row of $\mathbf{U}^{-1}$). Hence, in order to obtain $P_{\mathrm{ssDNA}}(t)$ for this case, one has to find the eigenvalues and eigenvectors of $\mathbf{T}$ and then evaluate this sum for each time point.\\    
To ensure meaningful sampling in order to fit a single-exponential in analogy to how it is done in experiments, we select a time range that avoids the trivial limits of \( P_{\mathrm{ssDNA}}(t) \to 0 \) and \( P_{\mathrm{ssDNA}}(t) \to 1 \) for forces below \(20~\mathrm{pN}\), our range of interest. 
Since external forces typically accelerate dissociation relative to spontaneous thermal melting, we set the end time \( t_e \) to be on the order of the thermal dissociation time of the 11bp duplex at room temperature
\begin{equation*}
    t_e = \frac{1}{k_{ot}} \left( \frac{k_c(F=0)}{k_o} \right)^{N_{bp}-1}
    = \frac{e^{-\left[\Delta G_{\mathrm{\text{non}}}(25^\circ \text{C}) + (N_{bp}-1)\Delta G_{s}(25^\circ \text{C})\right]/k_B T}}{k_a}=\frac{e^{-\Delta G_{\text{duplex}}(25^\circ \text{C})/k_BT}}{k_a}
    \approx \frac{10^6}{k_a}.
\end{equation*}
Once this characteristic timescale is established, Eq.~(\ref{equ:p_ss}) is evaluated at 1000 uniformly spaced time points within the interval \([0,\,t_e]\) (in units of \(1/k_a\)), see this reference~\autocite{hussein_2026_20597758} for the data and source code. This sampling was sufficient to capture the expected single-exponential behavior of \( P_{\mathrm{ssDNA}}(t) \), consistent with observations from comparable experimental studies.

\printbibliography

\end{document}